\documentclass[cmbright,fleqn,referee]{envauth}
\def\bSig\mathbf{\Sigma}

\received{00 Month 2018}
\revised{00 Month 2018}
\accepted{00 Month 2018}

\usepackage{natbib}
\usepackage{eqnarray}
\usepackage{url}
\usepackage{hyperref}
\usepackage{amsmath,amssymb,amscd,graphicx}

\usepackage{caption}
\usepackage{subcaption}
\usepackage{epstopdf}
\newtheorem{definition}{Definition}
\usepackage{booktabs}

\usepackage{multirow}
\usepackage{amsbsy}
\usepackage{gensymb}

\runninghead{Marwah Soliman, Vyacheslav Lyubchich and Yulia R. Gel}{Ensemble Forecasting of the Zika Space-Time Spread with Topological Data Analysis}

\begin{document}
	
	\title {Ensemble Forecasting of the Zika Space-Time Spread with Topological Data Analysis}
	\author{Marwah Soliman\affil{1}, Vyacheslav Lyubchich\affil{2} and Yulia R. Gel\affil{1}\corrauth}
	\corraddr{Yulia R. Gel , Department of Mathematical Sciences, University of Texas at Dallas, 800 West Campbell Road, Richardson, TX 75080, USA. E-mail: ygl@utdallas.edu}
	\address{\affilnum{1}Department of Mathematical Sciences, University of Texas at Dallas,
		Richardson, USA\\
		\affilnum{2}Chesapeake Biological Laboratory, University of Maryland Center for Environmental Science, Solomons, MD, USA}

\begin{abstract}
As per the records of the World Health Organization, the first formally reported incidence of Zika virus occurred in Brazil in May 2015. 
The disease then rapidly spread to other countries in Americas and East Asia, affecting more than 1,000,000 people.  
Zika virus is primarily transmitted through bites of infected mosquitoes of the species Aedes (\textit{Aedes aegypti} and \textit{Aedes albopictus}). The abundance of mosquitoes and, as a result, the prevalence of Zika virus infections are common in areas which have high precipitation, high temperature, and high population density. 
Nonlinear spatio-temporal dependency of such data and lack of historical public health records
make prediction of the virus spread particularly challenging. In this paper we enhance Zika forecasting by introducing the concepts of topological data analysis and, specifically, persistent homology of atmospheric variables, into the virus spread modeling. The key rationale is that topological summaries allow for capturing higher-order dependencies among atmospheric variables that otherwise might be unassessable via conventional spatio-temporal modelling approaches  based on geographical proximity assessed via Euclidean distance. 
We introduce a new concept of cumulative Betti numbers and then integrate the cumulative Betti numbers as topological descriptors into three predictive machine learning models: random forest, generalized boosted regression, and deep neural network. 
Furthermore, to better quantify for various sources of 
uncertainties, we
combine the resulting individual model forecasts into an ensemble of the Zika spread predictions using Bayesian model averaging. The proposed methodology is illustrated in application to forecasting of the Zika space-time spread in Brazil in the year 2018.

\end{abstract}

\maketitle

\section{Introduction}

The Zika virus is a flavivirus belonging to the \textit{Flaviviridae} family which also includes yellow fever, dengue, Japanese encephalitis, and West Nile viruses \citep{goeijenbier2016zika}.  The earliest known occurrence of Zika was identification of the virus in the serum of a rhesus monkey in 1947 in Uganda \citep{dick1952zika,dick1952zikaII}.
In 2015 Zika virus was detected in Brazil, with an estimate of  1.3 million infection cases, and rapidly was transmitted to other countries in North and South Americas, as well as East Asia~\citep{heukelbach2016zika, malone2016zika}. Symptoms of Zika virus infection among humans include headache, fever, reddened eyes, rashes, muscle and joint pain. Furthermore, Zika virus can cause severe birth defects when transmitted from a pregnant woman to her fetus. Given a rapid spread of the infection, the World Health Organization declared Zika an epidemic disease from 2015 to 
2016~\citep{WinNT4}.

Zika is primarily transmitted through the bite of infected \textit{Aedes aegypti} and \textit{Aedes albopictus} \citep{WinNT3}.
As a result, the spreading of Zika virus is significantly accelerated by weather conditions favoring the abundance of the Aedes mosquitoes, for example, in temperate climate zones with high humidity and rainfall \citep{tjaden2013extrinsic, rees2018environmental, munoz2017could}. In addition, the virus can be transmitted from mother to her fetus, also via sexual contacts, blood transfusion, and organ transplantation~\citep{WinNT4}.

Modeling and forecasting Zika spread continues to attract an ever increasing attention, and there exist three general methodological directions rooted in mathematics, statistics, and machine learning~\citep{ferraris2019zika}. Mathematical approaches include vector borne compartmental and mechanistic transmission models~\citep[see, e.g., overviews by][and references therein]{manore2016mathematical, Caminade119, suparit2018mathematical}. In turn, statistical methods tend to be largely based on the Box--Jenkins family of models with various exogenous predictors, such as atmospheric variables and data from HealthMap digital surveillance system, Google trends, and Twitter microblogs~\citep[see, e.g.,][and references therein]{mcgough2017forecasting, teng2017dynamic, castro2018implications}. Finally, machine learning approaches to modeling spread of Zika virus include random forest (RF), boosted regression (BR), and deep feed-forward neural network (DFFN) models~\citep{soliman2019complementing, seo2018decoding}. As recently shown by~\cite{soliman2019complementing}, the DFFN models tend to deliver more competitive predictive performance   
than RF and BR. Remarkably, while spatio-temporal epidemiology of (re)emerging infectious diseases has been extensively studied in the environmetrics community before~\citep[see, for instance,][and references therein]{nobre2005spatio,loh2011k, self2018large} and despite the increasing popularity of DL tools~\citep{mcdermott2019deep} in the spatio-temporal environmetrics applications, utility of DL approaches in infectious epidemiology remains largely unexplored.

Furthermore, in this paper we bring the tools of topological data analysis (TDA) 
to spatio-temporal modeling \citep{diggle2005point,torabi2013spatio,nobre2005spatio,jang2007comparison,ugarte2009evaluating}, and prediction of the Zika spread. TDA is an emerging methodology at the interface of computational topology, statistics, and machine learning that aims to enhance our understanding on the role of the underlying data shape in dynamics of the data generating process---in our case, the spatio-temporal process of the Zika spread which exhibits a complex nonstationary dependence structure.
Since areas with higher quantities of rainfall are principally associated with greater abundance of mosquitoes transmitting Zika~\citep{munoz2017could}, 
our key idea is to employ TDA to explore the structure and to extract information about the shape of the temperature and precipitation data that may be useful for predicting the spread of Zika infection.

While TDA is found to exhibit high utility in many fields, from genetics to finance to power systems \citep{gidea2018topological,saggar2018towards, li2020hybrid}, 
application of TDA within epidemiology~\citep{costa2014topological, lo2018modeling} and even more generally environmetrics~\citep{islambekov2019unsupervised, islambekov2019harnessing} still remains very limited. In particular, TDA and, specifically, persistent homology have been employed by~\cite{costa2014topological}
for analysis of influenza-like illness 
during the 2008--2013 flu seasons in Portugal and Italy.
Most recently, \cite{lo2018modeling} show explanatory power of TDA for analysis of Zika spread. Particularly, \cite{lo2018modeling} use persistent homology factors of the \textit{Aedes aegypti} mosquito occurrence locations as regressors within a linear model and show a high utility of topological features within a spatial cross-validation framework. The findings of~\cite{lo2018modeling} demonstrate that the model with topological features as regressors yields higher coefficient of determination ($R^2$) and lower cross-validation mean squared error than the benchmark linear model without TDA features. 

Our method advances the approach of~\cite{lo2018modeling} in multiple ways. In contrast to~\cite{lo2018modeling}, who do not consider prediction of Zika incidences over time, our goal is to assess utility of TDA in forecasting {\it future} spread of Zika, which is the key towards the outbreak emergency preparedness and response. 
Since mosquito occurrence locations may vary over time, any Zika prediction model based on topological features of mosquito locations also requires forecasts of the mosquito occurrences. Such forecasts are not readily available and may be highly non-trivial, especially in areas with heterogeneous landscapes as Brazil.

In contrast to~\cite{lo2018modeling}, we consider TDA in application to shape analysis of temperature and precipitation rather than mosquito data. Such approach allows us to systematically incorporate the future weather and climate forecasts from national weather services and derived topological features of such forecasts into Zika prediction models. Furthermore, we evaluate predictive utility of TDA based on multiple statistical and machine learning models, such as random forest, generalized boosted regression, and DFFN. 
We introduce a new topological summary concept {\it cumulative Betti number}, which is used as a predictor of future Zika dynamics and found to deliver a more stable performance than conventional topological characteristics.

Finally, to quantify multiple sources of uncertainty, we develop an adaptive ensemble of Zika forecasting models using Bayesian model averaging (BMA). We illustrate our proposed methodology in application to predicting Zika virus spread for all the 26 states of Brazil during the year 2018.

The key contributions of our paper can be summarized as follows:

\begin{itemize}
    \medskip
    \item 
    To the best of our knowledge, this is the first paper introducing topological data features as {\it predictors of future space-time spread} of infectious diseases. 
    
    \medskip
    \item
To increase stability of the topological summaries and associated derived forecasts, especially under the scenarios with limited sample sizes as in the case of many emerging climate-sensitive infectious diseases, we introduce the notion of {\it cumulative Betti numbers}. The proposed cumulative Betti numbers can be used in many other applications, beyond infectious epidemiology, which involve noisy data of moderate sample sizes, e.g., household travel networks and microgrids.

 \medskip
    \item
We validate utility of our proposed predictive approach across various nonparametric machine learning models, including deep neural networks, which allows for more systematic and objective assessment of predictive gains (if any) delivered by the proposed topological descriptors.
    
\end{itemize}

The remainder of the paper is organized in four major sections: data description, methodology for epidemiological forecasting and validation metrics, results, and discussion. In Section~\ref{DataDescription}, we provide information on the collected Zika rate, population density, and atmospheric data. We introduce our topological data analysis in Section~\ref{sec:TDA} and statistical and machine learning forecasting methodology in Section~\ref{machinemodels}. Section~\ref{Valid} lists the validation metrics, Section~\ref{Results} is devoted to validation of the proposed modeling approaches to prediction of Zika rate in Brazil. Finally, the paper is concluded with a discussion 
in Section~\ref{Discussion}.

\section{Data description}
\label{DataDescription}

Brazil's Ministry of Health publishes 
cumulative Zika rate on a weekly basis.
However, as with many other vector-borne  diseases, the official surveillance records for zika are often incomplete and noisy. For instance, 
for the year of 2018 the percentage of missing weeks is 28.8\%, or 15 weeks out of 52 weeks. In turn, 24 weeks out of 52 weeks are missing in the year of 2017, which is 46.15\%. Hence, we 
analyze monthly records 
of Zika virus rate per 100,000 individuals,
constructed from the cumulative weekly Zika rates published by Brazil's Ministry of Health~\citep{WinNT5}, in each of the 26 states of Brazil and each month during 2017 and 2018. 

Considering the strong association between Zika virus transmission and 
local environmental conditions 
\citep[e.g., see][]{tjaden2013extrinsic,
munoz2017could,rees2018environmental}, we also collected data for  precipitation and air temperature.
The weather station data are accessed for all states for years 2017 and 2018 through \cite{WinNT6}. The forecasted precipitation data for 2018 are obtained from~\cite{INMET}.
To ensure consistency in temporal resolutions of zika
and atmospheric data, all
daily atmospheric variables are aggregated to a monthly scale.

\section{Methodology for epidemiological forecasting} 
\label{StatMeth1}

\subsection{Topological data analysis}
\label{sec:TDA}

Topological data analysis (TDA) is a rapidly emerging methodology at the interface of computational topology, statistics, and machine learning which allows for a systematic multi-lens assessment of the underlying topology and geometry of the data generating process~\citep{zomorodian2005computing, carlsson2009topology, chazal2017introduction}. 
In this paper, we primarily employ tools of persistent homology (PH) within the TDA framework. The ultimate idea of PH is to quantify dynamics of topological properties exhibited by the data that live in Euclidean or abstract metric space, at various resolution scales. 
The PH approach is implemented in the two key steps: first, the underlying hidden topology of the observed data set is approximated using certain combinatorial objects, e.g., simplicial complexes, and then the evolution of these combinatorial objects is studied as the resolution scale varies. 

We start with providing a brief overview of the main relevant technical concepts.


\begin{definition}[Abstract simplicial complex]
		Let $P$ be a discrete set. Then,
		an abstract simplicial complex is 
		a collection $K$ of finite nonempty subsets of $P$ such that if $\sigma \in K$ and $\tau \subset \sigma$, then $\tau \in K$.	
		If $|\sigma|=k+1$, then $\sigma$ is called a $k$-simplex. 
\end{definition}

In case of an Euclidean space, $k$-simplex corresponds to a convex hull of $k+1$ vertices. Hence, a 0-simplex is a vertex, a 1-simplex is an edge, a 2-simplex is a triangle, and a 3-simplex is a tetrahedron. 

One of the most widely used choices for a simplicial complex within the TDA framework is a Vietoris--Rips complex which has gained its popularity due to its computational properties and tractability.

\begin{definition}[Vietoris--Rips complex] 
Let $(X, d)$ be a metric space, e.g., $\mathbb{R}^{m}$  and $P \subseteq X$ be a set of distinct points of $X$. 
Let $\epsilon>0$ be a scale. Then the Vietoris--Rips complex $V_{\epsilon}(P)$ is an abstract simplicial complex whose finite simplices $\sigma$ in $P$ have a diameter at most $\epsilon$, i.e.,
    $$V_{\epsilon}(P) = \{\sigma \subseteq P | d(u,v)\leqslant \epsilon,\; \forall u \neq v \in \sigma\}.$$


\end{definition}

That is, we form the {\it proximity graph} of $P$ by joining two points in $P$ whenever their pairwise distance is less than $\epsilon$.
We now consider a sequence of scales $\epsilon_1 < \epsilon_2 < \ldots < \epsilon_n$ and associated nested sequence of VR complexes called a Vietoris--Rips filtration $V_{\epsilon_1}(P) \subseteq V_{\epsilon_2}(P) \subseteq \ldots \subseteq V_{\epsilon_n}(P)$. As a result, we can study evolution of topological summaries, such as number of connected components, loops, etc., that appear and disappear with an increase of scale $\epsilon$. Persistent topological features, i.e., those with a longer lifespan over varying resolution $\epsilon_1 < \epsilon_2 < \ldots < \epsilon_n$ tend to be associated with the underling structural organization of the data generating process, while features with a shorter lifespan are likely to be a topological noise. 




\begin{figure}[h]
     \centering
     \begin{subfigure}[b]{0.2\textwidth}
         \centering
        \includegraphics[width=\textwidth]{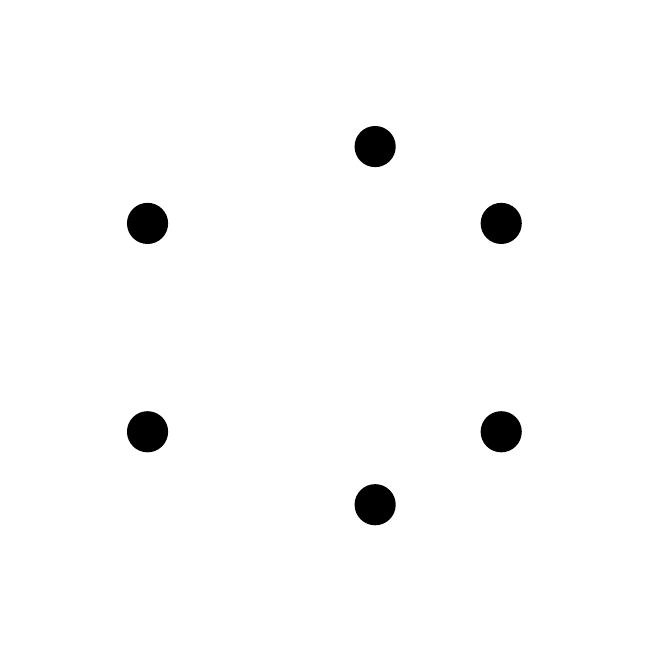}
        \caption{$\epsilon = 0$}
         \label{fig:epsilon0}
     \end{subfigure}
     \hfill
     \begin{subfigure}[b]{0.2\textwidth}
        \centering
         \includegraphics[width=\textwidth]{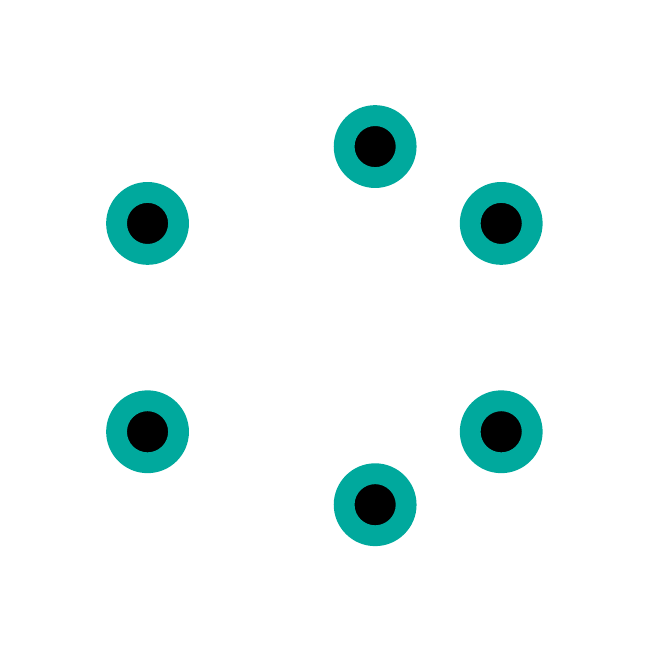}
         \caption{$\epsilon = 0.2$}
         \label{fig:epsilon1-2}
     \end{subfigure}
    \hfill
     \begin{subfigure}[b]{0.2\textwidth}
         \centering
         \includegraphics[width=\textwidth]{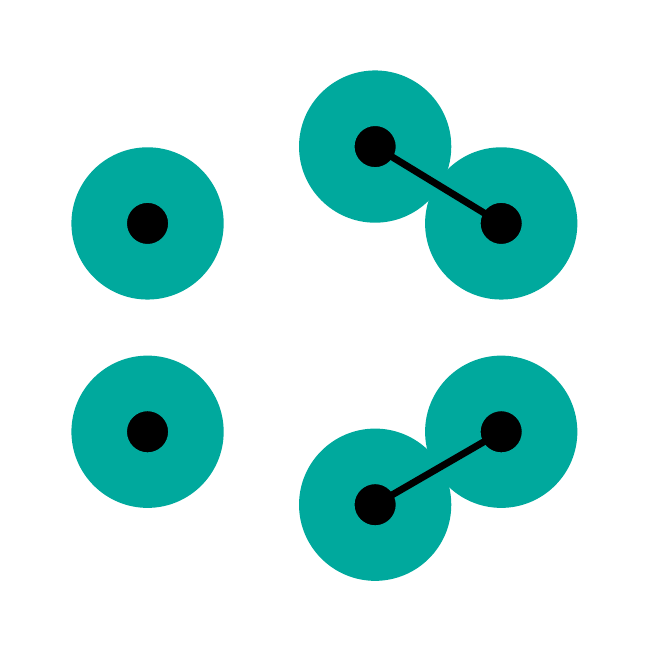}
         \caption{$\epsilon = 0.5$}
         \label{fig:epsilon1}
     \end{subfigure}
     \hfill
     \begin{subfigure}[b]{0.2\textwidth}
         \centering
         \includegraphics[width=\textwidth]{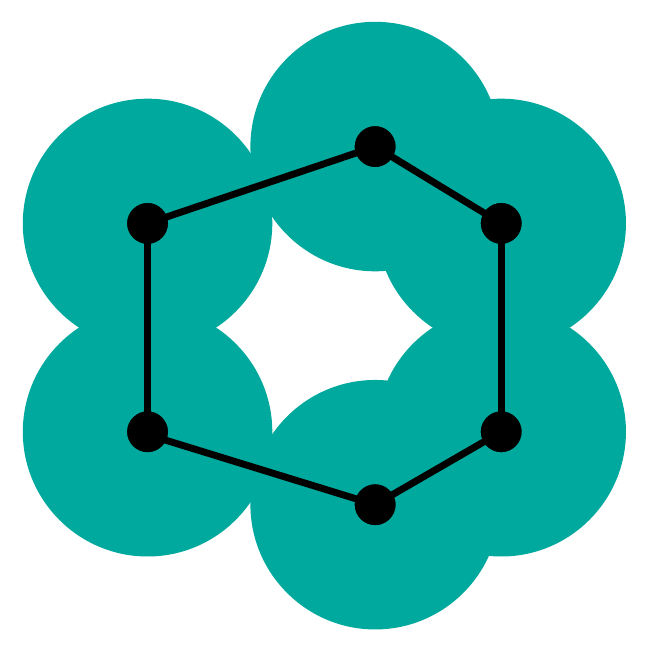}
         \caption{$\epsilon = 1$}
         \label{fig:epsilon15}
     \end{subfigure}
        \caption{\Large An illustration of Vietoris--Rips filtration with varying scales $\epsilon$. 
}
\label{fig:betti}
\end{figure}

Figure~\ref{fig:betti}
is an illustration of the Vietoris--Rips filtration process based on a toy example of six points. The filtration starts with a ball of  radius 0 ($\epsilon=0$) around each point (see Figure~\ref{fig:epsilon0}). As the scale $\epsilon$ increases to 0.2, 0.5, and 1 (see Figures~\ref{fig:epsilon1-2},~\ref{fig:epsilon1}, and ~\ref{fig:epsilon15}, respectively), the number of connected components decreases and new topological features, such as the loop, appear.

There exist multiple topological summaries to quantify evolution of topological features over the increasing scale $\epsilon$, e.g., barcode, persistent diagrams, persistent landscapes, and Betti numbers \citep[see the discussion in][and references therein]{chazal2017introduction}.
In this project we focus on utility analysis of Betti numbers.

\begin{definition}[Betti number]
The Betti-$k$ number $\beta_{k}$ is the rank of the $k$-th homology group. That is, $\beta_{k}(\epsilon)$ is the number of $k$-dimensional simplicial complex features for a given scale $\epsilon$. 
\end{definition}

For a given scale $\varepsilon$ and a given abstract simplicial complex, e.g.,  the  Vietoris--Rips complex, constructed from the observed point cloud under scale $\varepsilon$, Betti numbers are simply counts of particular topological features in this abstract simplicial complex. For instance, $\beta_{0}$ is the count of connected components in the  Vietoris--Rips complex; while $\beta_{1}$ and $\beta_{2}$ are the numbers of holes and voids in the  Vietoris--Rips complex, respectively.

Furthermore, we introduce the concept of {\it cumulative Betti numbers} which, as we find, tend to deliver more stable predictive performance for forecasting the Zika spread. 

\begin{definition}[Cumulative Betti numbers] Over a sequence of scales $\epsilon_1<\ldots<\epsilon_n$, cumulative Betti number $\tilde{\beta}_{k}(\epsilon_m)$, $m\leqslant n$, is defined as the sum of Betti numbers $\beta_{k}(\epsilon_i)$, $i=1,\ldots, m$. 
That is,
\begin{eqnarray}
\label{cumul_Betti}
\tilde{\beta}_{k}(\epsilon_m)=\sum_{i=1}^m \beta_{k}(\epsilon_i), \quad m\leqslant n.
\end{eqnarray}
\end{definition}

In this paper, due to the limited data records, we use only the 
$\tilde{\beta}_{0}$ numbers as topological descriptors, that is, the cumulative number of connected components. We hypothesize that higher predictive performance delivered by cumulative Betti numbers may be explained by higher robustness of $\tilde{\beta}_{0}$ to scale sequence selection under uncertainty due to low sample sizes. We used the function \texttt{ripsDiag} in R package \emph{TDA} \citep{TDA} to find the cumulative Betti~0 numbers.

\begin{figure}[ht]
     \centering
     \begin{subfigure}{0.475\textwidth}
         \centering
         \includegraphics[width=0.85\textwidth, viewport = -10 10 400 272, clip]{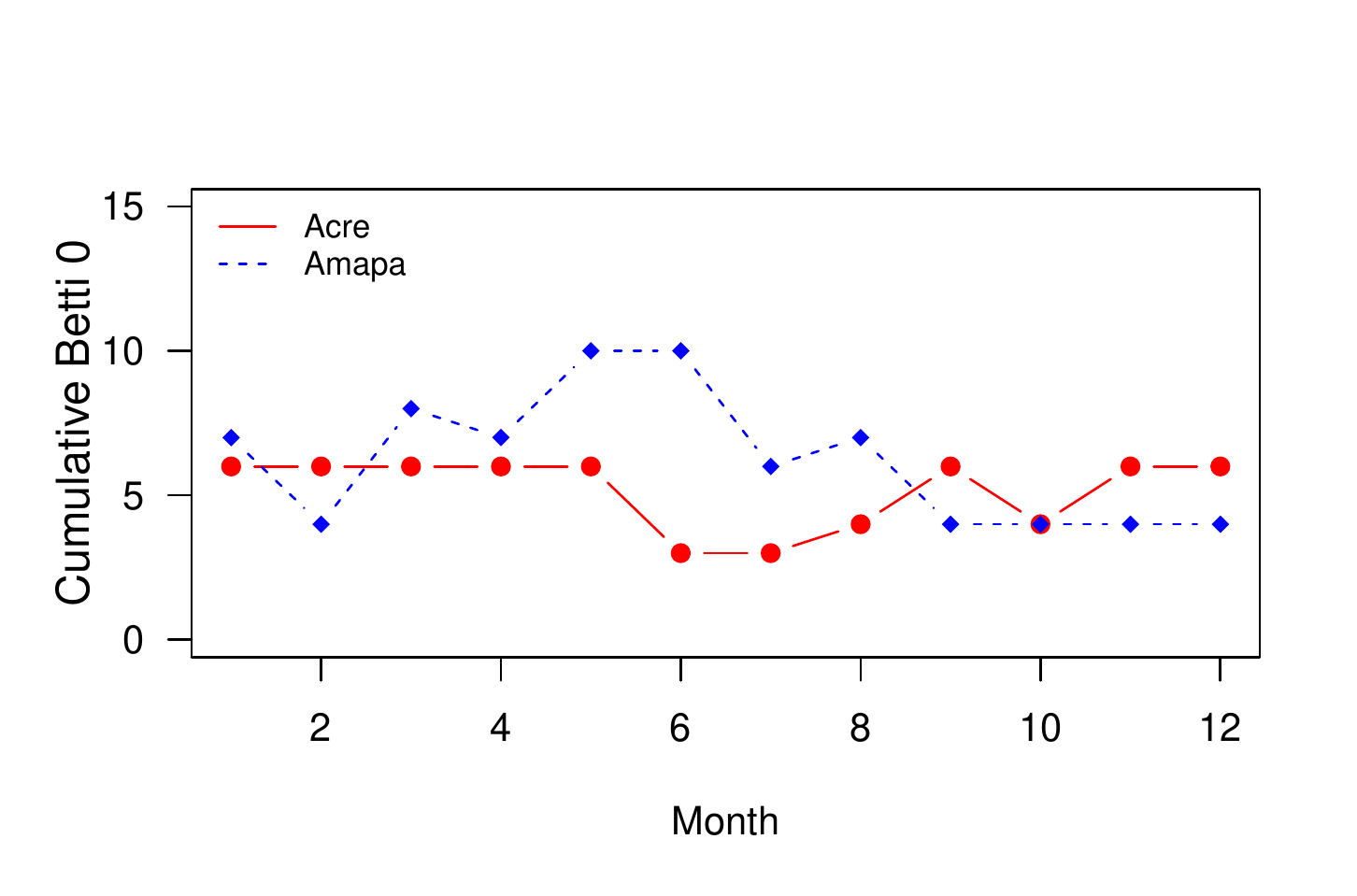}
         \caption{Precipitation, mm}
        \label{fig:pcp}
     \end{subfigure}
     \begin{subfigure}{0.475\textwidth}
         \centering
     \includegraphics[width=0.9\textwidth, viewport = -10 10 400 272, clip]{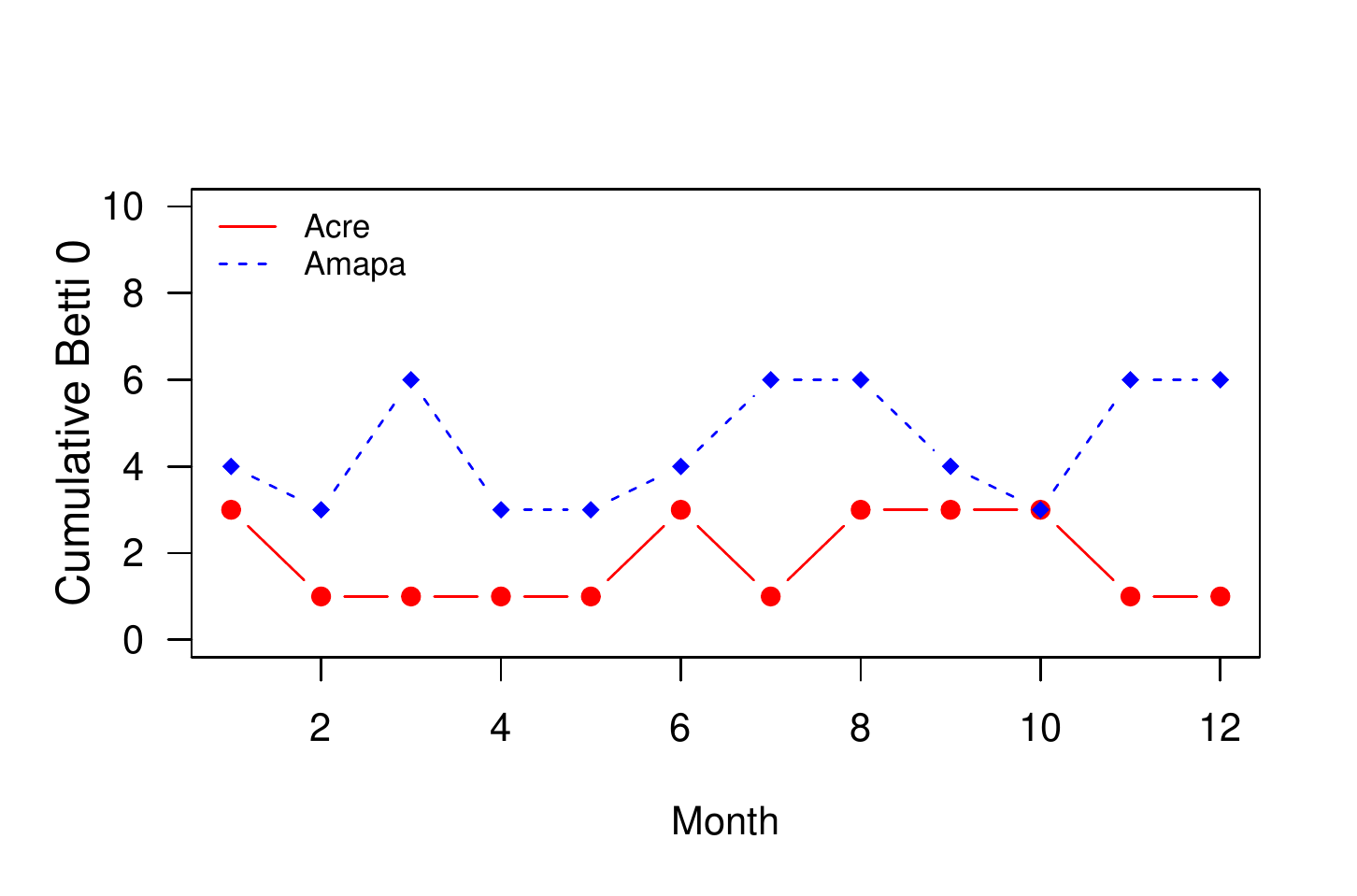}
         \caption{Temperature, {\degree}C}
          \label{fig:temp}
     \end{subfigure}
        \caption{\Large Cumulative Betti-0,  $\tilde{\beta}_{0}$, numbers for precipitation and temperature in the states Acre and Alagoas of Brazil in 2017.
        }   
        \label{fig:map4}
\end{figure}

Figure~\ref{fig:map4} depicts cumulative Betti-0,  $\tilde{\beta}_{0}$, numbers based on precipitation amounts and temperature in the states of Acre and Alagoas, Brazil.

To extract the topological characteristics for precipitation ($X_{i,j}^{3}$) and temperature ($X_{i,j}^{4}$), we use the following algorithm.

Let $M_{j}$ be the number of weather stations in each state $j$ and $X_{i,j,k}$ be the corresponding weather observations in month $i$ ($i=1,\ldots,12$; $j =1,\ldots,26$, and $k=1,\ldots,M_{j}$). (Here we suppress the superscripts for the sake of notations.) 
Consider a point cloud 
$\{X_{i,j,1}, X_{i,j,2},\ldots, X_{i,j,M_j}\}$ for each month $i$ and state $j$. 
We now measure  (dis)similarity (in terms of the Euclidean distance) among recorded atmospheric variables in the $i$-th month across all weather stations $k=1,\ldots, M_j$ in the $j$-th state. For instance, in a case of temperature ($X_{i,j}^{4}$), 
we set up $\epsilon$ (in degrees Celsius) and obtain a distance graph by connecting two weather stations $l$ and $m$ only if their recorded
temperature observation 
differ at most $\epsilon$ in degrees Celsius,
i.e., $|X_{i,j,l}-X_{i,j,m}|\leq \epsilon$. We now count a number of
topological features (e.g.,
a number of connected components, or Betti~0 $\beta_{0}(\epsilon)$) for a given scale $\epsilon$. 
Then we increase scale $\epsilon$ and repeat the procedure.
Summing over consider scales (see~(\ref{cumul_Betti})) yields a cumulative Betti number for the $i$-th month in the $j$-th state 
($i=1,\ldots,12$; $j =1,\ldots,26$).
The idea is that topological summaries allow for capturing higher-order dependencies among atmospheric variables that are unassessable via conventional spatio-temporal modelling approaches based on geographical proximity.

\subsection{Models}
\label{machinemodels}

We employ three statistical and machine learning models to predict the Zika activity at the state level, namely, random forest (RF), boosted regression (BR), and deep feed-forward neural networks (DFFN).   
Using Bayesian model averaging (BMA), we then develop a multi-model ensemble of the Zika forecasts.  
We train all the models on the year 2017 data (training set), then evaluate their out-of-sample forecasts using the 2018 data (testing set). 
Below we provide an overview of the considered modeling approaches and the forecast validation metrics. 

Let $\mathbf{Y}_{i,j}$ be the Zika rate in month $i$ in state $j$ ($i=1,\dots,12$ and $j=1,\dots,26$) and
$\mathbf{X}=\{\mathbf{X}_{i,j}^{1},\ldots, \mathbf{X}_{i,j}^{7}\}$ be the corresponding regressors. 
In each of the models, $\mathbf{X}_{i,j}^{1}$ is the precipitation, $\mathbf{X}_{i,j}^{2}$ is the temperature,  $\mathbf{X}_{i,j}^{3}$ is the cumulative Betti-0 
for the precipitation, $\mathbf{X}_{i,j}^{4}$ is 
the cumulative Betti-0 
for the temperature, $\mathbf{X}_{i,j}^{5}$ is the categorical variable representing month (January, \dots, December), $\mathbf{X}_{i,j}^{6} = \mathbf{X}_{i-1,j}^{1}$ is the precipitation lagged by one month, and $\mathbf{X}_{i,j}^{7} = \mathbf{X}_{i-1,j}^{2}$ is the temperature lagged by one month. We added the lagged variables ($\mathbf{X}_{i,j}^{6}$ and $\mathbf{X}_{i,j}^{7}$) to account for delay effects.

\paragraph{Random forest (RF)}

The random forest is one of the most popular models in machine learning and statistics, with an idea to combine several individual decision tree models into an additive multi-model ensemble~\citep{breiman2001random}:
$$g(x)=\sum_{k\in\mathbb{Z}^+ }f_k(x), $$
where $f_k(x)$ is an individual decision or regression tree. The two fundamental ideas behind the RF approach are i)  for constructing individual trees $f_k$, observations from the training set are randomly sampled with replacement (i.e., bootstrapped), and ii)  
each individual split in a tree is based on a random subset of variables that are used in the model. These concepts of RF allow to reduce the effect of overfitting and to optimize a bias-variance trade-off.
We used the function \texttt{randomForest} in R package \emph{randomForest}~\citep{RF} to train the random forest model.

\paragraph{Generalized boosted regression model} 


The model is built on two techniques: decision tree algorithms and boosting methods~\citep{Breiman, friedman2001greedy, friedman2002stochastic}. Generalized boosted models iteratively fit many decision trees to improve the accuracy of the model. 

Let $J$ be the number of terminal nodes in a  regression tree. The tree partitions the input space $X$ into $J$ disjoint regions ${\displaystyle R_{1},\ldots,R_{J}}$. Each regression tree model itself takes an additive form
$$ h(x;\{b_{j},R_{j}\}_{1}^{J})=\sum_{j=1}^{J}b_{j}I(x\in R_{j}),$$
where $b_{j}$ is the value predicted in the region $R_{j}$, $x$ is the input datum, and $I(\cdot)$ is the indicator function.
Then, the update at the $m$-th iteration, $m\in \mathbb{Z}^{+},$ is
$$ F_{m}(x)= F_{m-1}(x)+ \rho_{m}\sum_{j=1}^{J}b_{jm}I(x\in R_{jm}),$$
where $\rho_{m}$ is a scaling factor and the solution to the ``line search'';
$\{b_{jm}\}_{j=1}^{J}$ are the corresponding least-squares coefficients, and $\{R_{jm}\}_{j=1}^{J}$ are the regions defined by the terminal nodes of the tree at the $m$-th iteration. 

Let $\gamma_{jm} = b_{jm} \rho_{m}$.
Note that a separate choice for optimal value $\gamma_{jm}$ is proposed by \citet{friedman2001greedy} in each tree region, opposed to a single $\gamma_{m}$ in an entire tree. Update rule for model now becomes:
\begin{align}
\hspace{10em}
F_{m}(x)&=F_{m-1}(x)+\sum _{j=1}^{J}\gamma_{jm}I(x\in R_{jm}),\label{F}  \\
   \gamma_{jm}&=\underset {\gamma }{\operatorname {arg\,min} }\sum _{x_{i}\in R_{jm}}L(y_{i},F_{m-1}(x_{i})+\gamma ), \label{gamma_up}
\end{align}
where~\eqref{gamma_up} is just the optimal constant update in each terminal node region, based on the loss function $L$, given the current approximation $F_{m-1}(x_i)$. 


Additional boosting algorithm details are available in \citet{ridgeway2007generalized} and \citet{friedman2001greedy}. For the boosted regression, we used the function \texttt{gbm} in R package \emph{gbm}~\citep{gbm} with the number of trees equal to 100 and the interaction depth of 3, chosen through a cross-validation.


\paragraph{Deep feed-forward neural network (DFFN)}
The DFFN algorithm \citep{hagan1994training,riedmiller1993direct,zhang2007hybrid} is performed as follows: 
input data move forward from input to hidden layers to output, and at each of those moves the data are non-linearly transformed (using so-called {\it activation function}) and re-weighted. When reaching the output layer, the error is calculated, based on a selected cost function, which reflects how far the DFFN output is with respect to the actual data (i.e., Zika rate) in the training set. 





To illustrate the DFFN algorithm, let $w^l_{jk}$ be the weight connecting the $k$-th neuron in the $(l-1)$th layer to the $j$-th neuron in the $l$-th layer, and let $\sigma(\cdot)$ be the activation function (we use sigmoid function), where $j,k,l\in\mathbb{Z}^+$. The activation values at each layer are calculated in the forward procedure as
$$ a_j^l = \sigma\left(\sum_{k}w^l_{jk}a_k^{l-1}+b^l_{j}\right),$$
where $b^l_{j}$ is the bias coefficient, analogous to the intercept term in regression models. After calculating the total error at the output layer, 
a partial derivative of the cost function with respect to each weight and bias term is obtained.
We train DFFN using \texttt{h2o.deeplearning} function in R package \emph{h2o}~\citep{h2o}. The optimal structure for DFFN is selected through a cross-validation: with 2 hidden layers, 12 nodes in each layer, and learning rate~0.01.

\paragraph{Bayesian model averaging (BMA)}

We evaluate uncertainty of epidemiological forecasts by constructing a weighted multi-model ensemble. This application of BMA enables a combination of multiple models, with their respective weights assigned according to the models' most recent predictive performance~\citep{hoeting1999bayesian,raftery2005using,fragoso2018bayesian}. 

In this project, we define model weights within a BMA using root mean square error (RMSE) calculated for the training set of data:
$$ RMSE(s) =\sqrt{\frac{\sum_{i=1}^{12}\sum_{j=1}^{26} ({y}_{ij}-\tilde{y}_{ij}(s))^{2}}{n}},$$
where ${y}_{ij}$ 
is the observed data point (in our case, the observed Zika rate in $i$-th month of 2017 at the $j$-th state), $\tilde{y}_{ij}(s)$ is the corresponding estimate delivered by the $s$-th model ($s=1,\ldots, S$),
$n$ is the total size of the training data set, and $S$ is the number of models. Then, the resulting weight $\omega_s$ for the $s$-th model in the BMA ensemble of forecasts is computed as follows:
$$ \omega_s=\frac{1/RMSE(s)}{\sum_{i=1}^{3}1/RMSE(i)}.  
$$

Now, let $Pred(s)$ ($s=1, \ldots, S$) denote an out-of-sample forecast produced by the $s$-th model (in our case, the Zika rate forecast of the $s$-th model for the $j$-th state and $i$-th month in 2018). Then the multi-model BMA forecast for that state is defined as
$$ Pred_{RMSE}= {\sum_{s=1}^{S} \omega_s Pred(s)}.
 $$

We set $s = 1$ to represent the random forest model, $s = 2$ to represent generalized boosted regression, and $s = 3$ to correspond to the deep feed-forward neural network.

\subsection{Validation metrics}
\label{Valid}
We employ two standard statistical measures of accuracy, that is, root mean square error (RMSE) and mean absolute error (MAE)
\citep{bluman2009elementary, peck2015introduction, kim2019weekly}. Let $y_{i,j}$ and $\hat{y}_{i,j}$ be the observed and predicted Zika rates for the $i$-th month in $j$-th state in the testing data set. Then,  
$$ RMSE = \sqrt{\frac{\sum_{i=1}^{12}\sum_{j=1}^{26} (\hat{y}_{i,j}-y_{i,j})^{2}}{N}}, $$
$$ MAE  = \frac{1}{N}\sum_{i=1}^{12}\sum_{j=1}^{26} \left|{{\hat{y}_{i,j}-y_{i,j}}}\right|, $$
where $N$ is the size of the testing data set. 
In addition, correlation between observed and forecasted values is measured through the Pearson correlation coefficient
$$ r(y,\hat{y})={\frac {\sum _{i=1}^{12}\sum_{j=1}^{26}(\hat{y}_{i,j}-{\bar{\hat{y}}})(y_{i,j}-{\bar {y}})}{{\sqrt {\sum _{i=1}^{12}\sum_{j=1}^{26}(\hat{y}_{i,j}-{\bar{\hat{y}}})^{2}}}{\sqrt {\sum _{i=1}^{12}\sum_{j=1}^{26}(y_{i,j}-{\bar {y}})^{2}}}}}, $$
where $\bar{y}$ denotes average Zika rate across all states of Brazil in year 2018, and $\bar{\hat{y}}$ is the average of out-of-sample predictions across all the states for the corresponding period.
 
Lower RMSE, MAE, and higher Pearson correlation coefficient imply better forecasting performance.

\section{Results}
\label{Results}

We train our models based on the data observed in the year of 2017
and use the 2018 data for verification. In particular,
let $\mathbf{Y}_{i,j}$ be a vector of Zika rates in the $i$-th month and the $j$-th state of year 2017, and let $\mathbf{X}$ be a matrix of regressors (i.e., design matrix)
in the year of 2017.
We now compare the performance of the models with persistent features (i.e.,
with the full set of inputs $X_{i,j}^{1},\ldots,X_{i,j}^{7}$)
against the models without persistent features (i.e., including only $X_{i,j}^{1}$, $X_{i,j}^{2}$, and $X_{i,j}^{5}$ 
as the inputs).
As Table~\ref{tableRMSE_Welch}
demonstrates, models with persistent features tend to perform better than models without persistent features.

To address whether predictive power of models with topological features is significantly different from models without topological features, we perform Welch's $t$-test~\citep{welch1947generalization, starnes2010practice} on the absolute and squared errors delivered by models with and without topological predictors, under the null hypothesis that topological predictors yield no predictive gain.
In addition, we assess whether the deep learning model (i.e., a deep feed-forward network) yields a significantly different predictive gain comparing to random forest and boosted regression.

\begin{table}[ht]
\caption{Performance summary for out-of-sample forecasts in terms of
MSE, MAE and  Welch's $t$-test $p$-values constructed for MSEs and MAEs of the models with and without topological features
}
\label{tableRMSE_Welch}
\centering
\resizebox{\columnwidth}{!}{%
        \begin{tabular}{lrrrrrrrrr}
        \hline
       & \multicolumn{3}{c}{With persistent features} & \multicolumn{3}{c}{Without persistent features}
       & \multicolumn{1}{c}{MSE}
       & \multicolumn{1}{c}{MAE}
       \\ \cline{2-7}
			Method & RMSE  & MAE & $r(y,\hat{y})$ & RMSE & MAE & $r(y,\hat{y})$ & $p$-value & $p$-value \\
	\hline 
	Random forest & 11.351& 8.262 &0.391 &12.134& 8.667 &0.307& 0.348 & 0.396 \\ 
	Boosted regression & 13.238 & 9.154& 0.363 &15.044 & 10.091 &0.286 & 0.093 & 0.262\\ 
	DFFN &8.934&6.361&  0.425 & 10.953& 8.084 &0.347 & 0.001& 0.007\\ 
	BMA-RMSE
	&6.721&4.265 & 0.419 & 8.380 &7.854&0.397 & $<0.001$ & $< 0.002$\\ 
	\hline 
\end{tabular}%
}
\end{table}

\begin{table}[ht]
\caption{
Welch's $t$-test $p$-values for  
comparison of predictive gains delivered by the 
deep feed-forward network (DFFN) vs.\ random forest and boosted regression
}
\label{tableRMSE_DL}
\centering
 \begin{tabular}{lrr}
        \hline
Model       & \multicolumn{1}{c}{With persistent features} & \multicolumn{1}{c}{Without persistent features}
       \\ 
 \hline
	Random forest & 0.002 & 0.277 \\ 
	Boosted regression & $<0.001$ & 0.203\\ 
	\hline 
	\end{tabular}
\end{table}

Table~\ref{tableRMSE_Welch} indicates that topological features lead to a highly statistically significant predictive gain, while incorporated into the DL DFFN model. However, while topological features result in lower predictive RMSEs for random forest and boosted regression, the predictive gains evaluated using Welch's $t$-test, appear to be not significant. In turn, Table~\ref{tableRMSE_DL} suggests that DFFN with persistent features 
delivers the most competitive predictive performance  among the three individual models, with Welch's $t$-test $p$-values of $<0.001$.

As expected, given the DFFN competitiveness, the BMA forecasting results are largely driven by the performance dynamics of DFFN and as such, topological descriptors are found to deliver highly statistically significant forecasting utility within the BMA framework (see Table~\ref{tableRMSE_Welch}).


For example, the incorporation of the persistent features lowers the RMSE of RF by 6.5\%, BR by 12.0\%, and DFFN by 18.4\%, while the RMSE of the BMA decreases by 19.8\%. Similar improvements are observed in terms of MAE, with the MAE of BMA decreasing by 45.7\%.



\begin{figure}[ht]
     \centering
     \begin{subfigure}[b]{0.25\textwidth}
         \centering
         \includegraphics[width=1.5\textwidth,scale = 1.1, viewport = 1 10 450 314, clip]{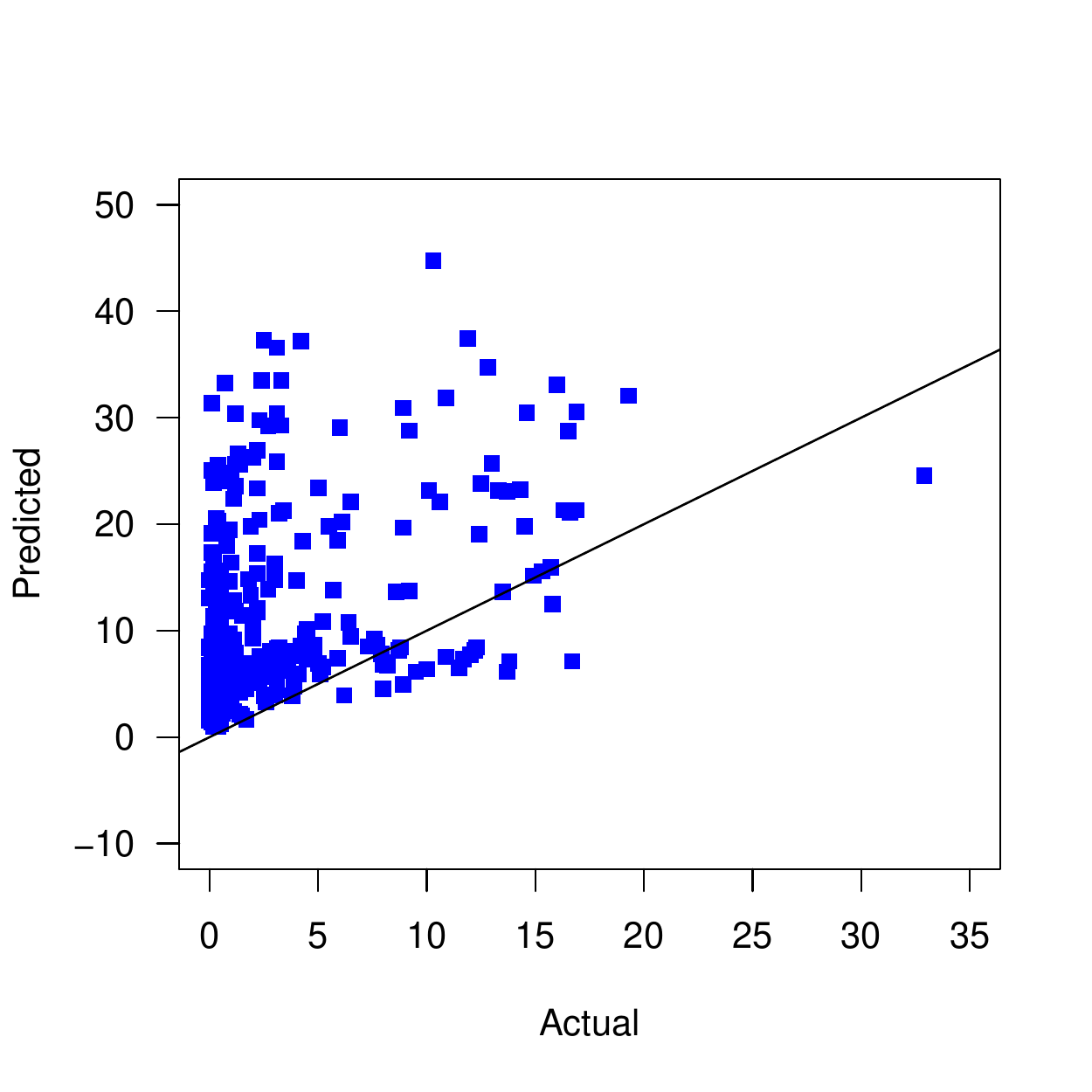}
         \caption{Random forest}
        \label{fig:RF1}
     \end{subfigure}
     \hspace{4em}
     \begin{subfigure}[b]{0.25\textwidth}
         \centering
         \includegraphics[width=1.5\textwidth,scale = 1.1, viewport = 1 10 430 314, clip]{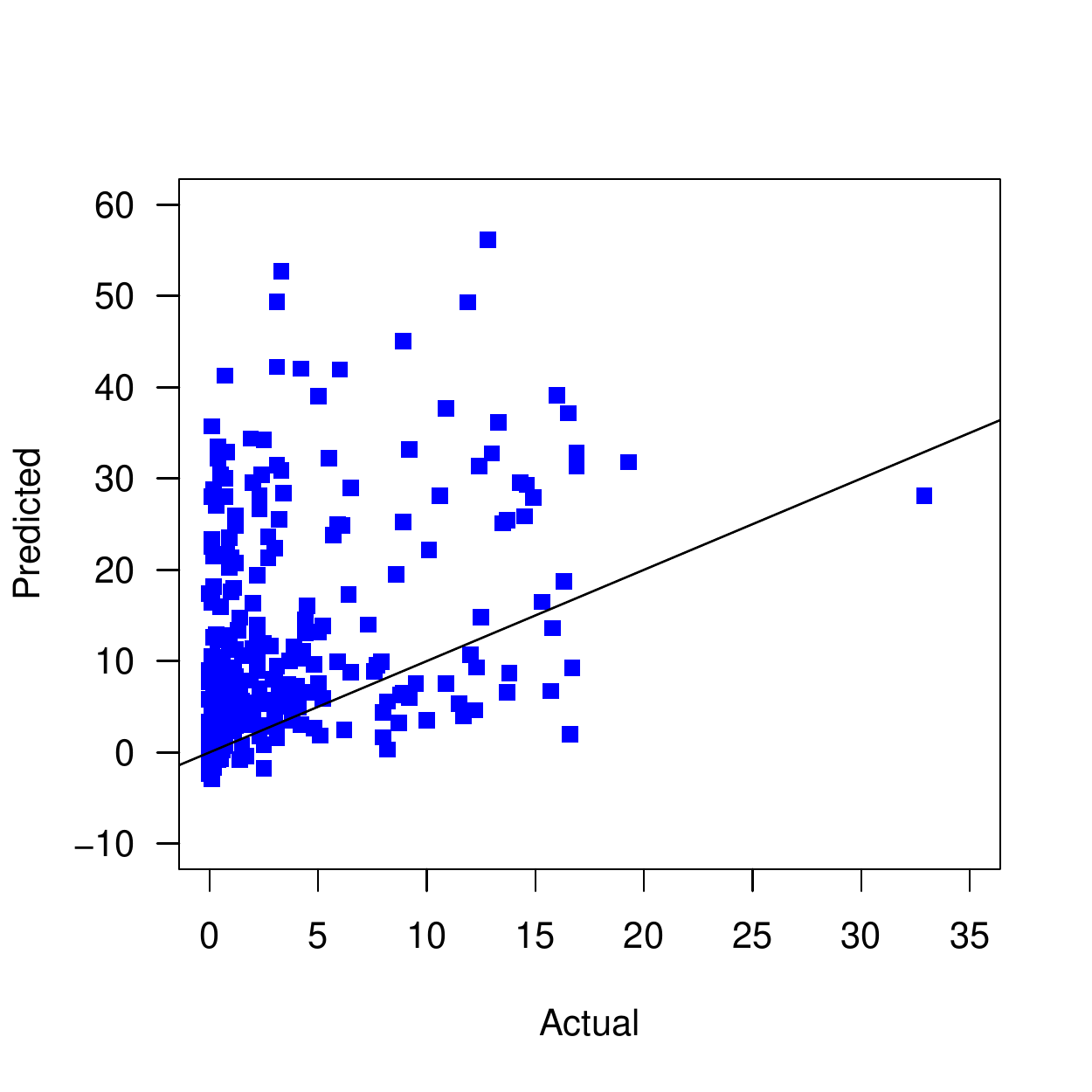}
         \caption{Boosted regression}
         \label{fig:predBR}
     \end{subfigure}
  \vskip\baselineskip
     \begin{subfigure}[b]{0.25\textwidth}
         \centering
         \includegraphics[width=1.5\textwidth,scale = 1.1, viewport = 1 10 430 314, clip]{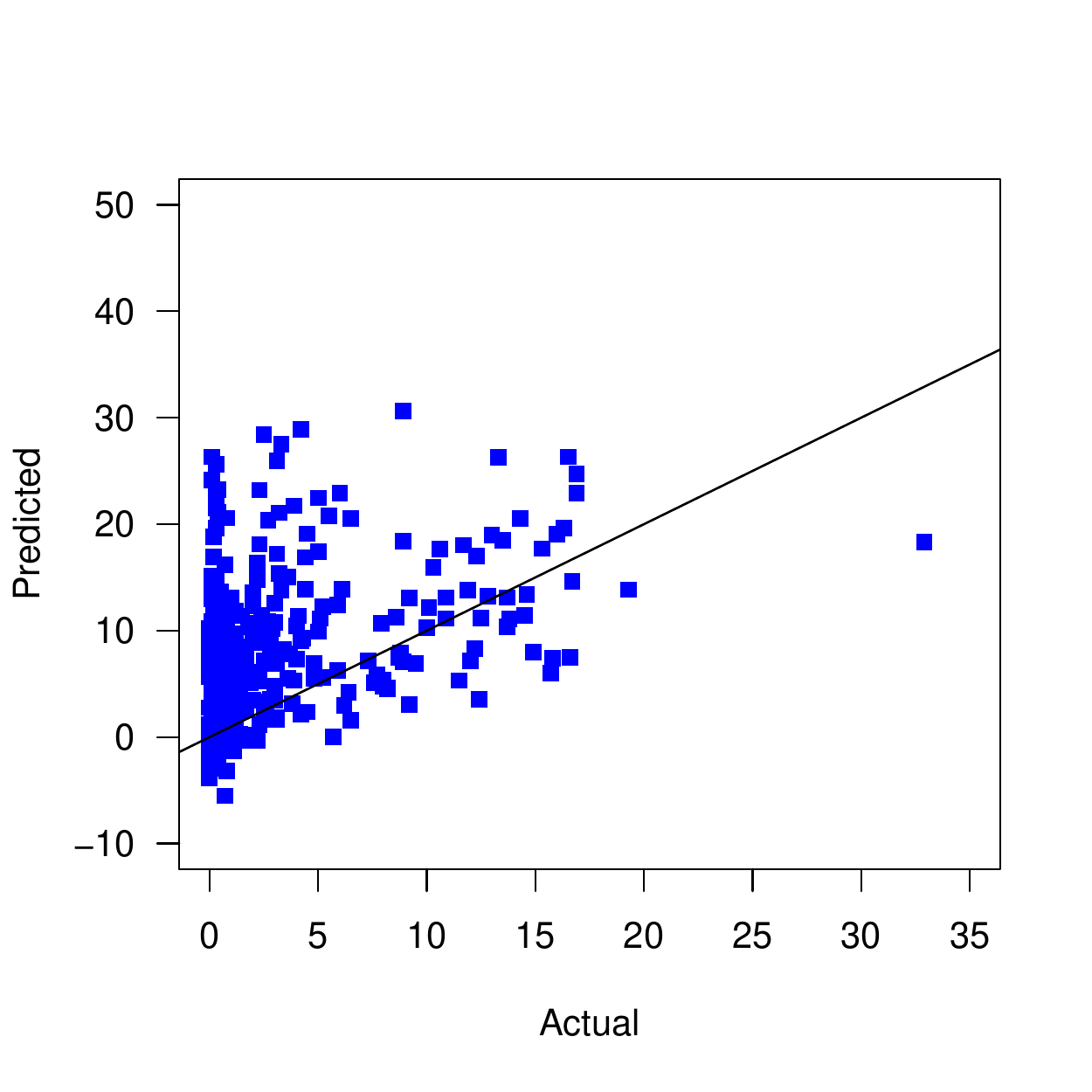}
         \caption{DFFN}
          \label{fig:predDFFN}
     \end{subfigure}
    \hspace{4em}
     \begin{subfigure}[b]{0.25\textwidth}
         \centering
         \includegraphics[width=1.5\textwidth,scale = 1.1, viewport = 1 10 430 314, clip]{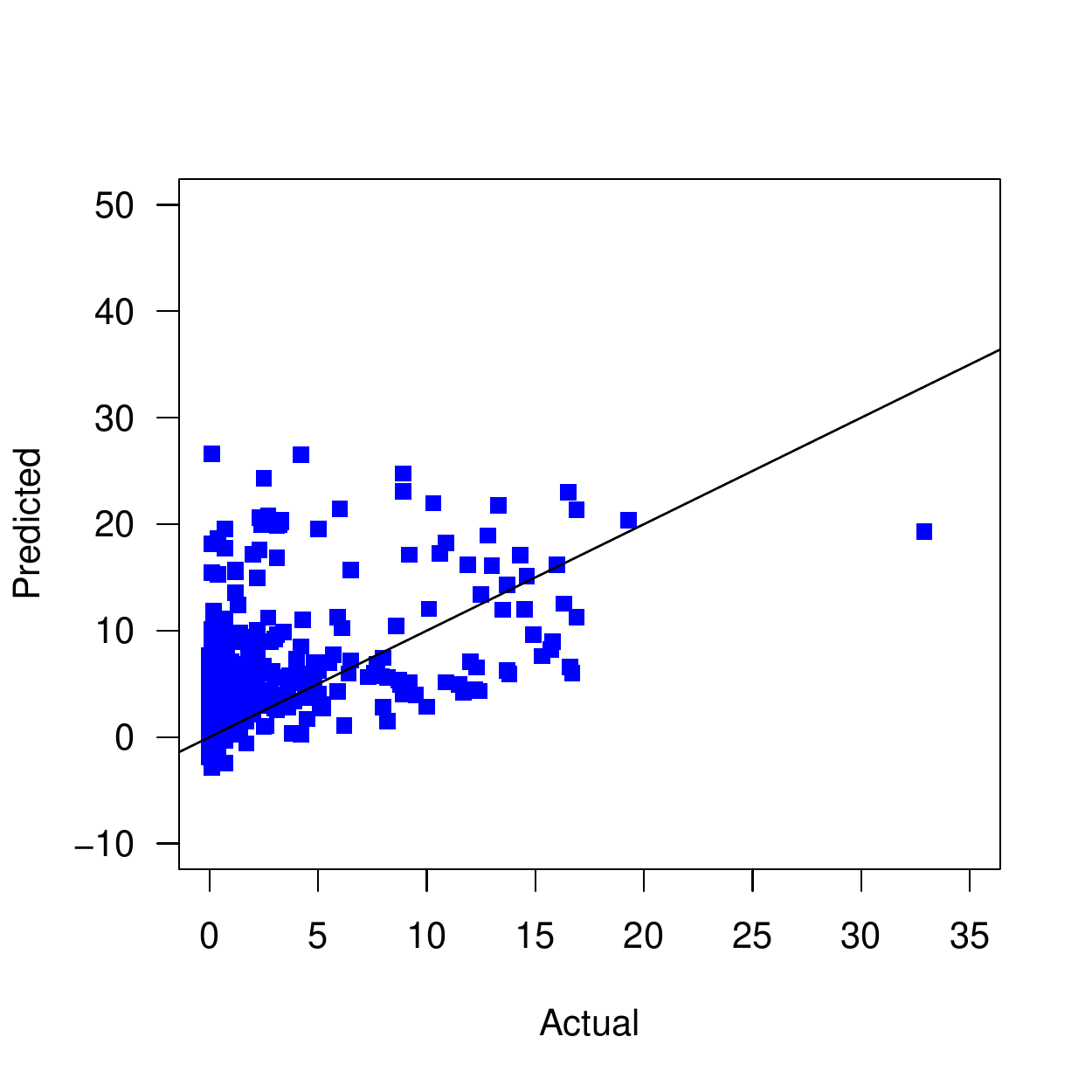}
         \caption{BMA-RMSE}
         \label{fig:predBMA}
     \end{subfigure}
        \caption{\Large Predicted values from each of the four models vs.\ the actual Zika rate in 2018. The black 45-degree lines represent the case of ideal prediction accuracy.
        }   
      \label{fig:prediction}
\end{figure}


Among the three individual models considered in this study, the tree-based models does not deliver the best forecasting performance. 
Figure~\ref{fig:RF1} shows the RF
predictions are grouped  above the 45-degree line, denoting over-prediction of the Zika rates; bias of RF is a well known issue (see technical details, e.g., in Chapter~15 by \citealp{Hastie:etal:2009}). The predictions from other models are less biased overall. Particularly, DFFN (Figure~\ref{fig:predDFFN}) and BMA (Figure~\ref{fig:predBMA}) show a better balance of over-prediction and under-prediction. However, the individual errors are larger (points in the graphs for BR are farther from the 45-degree line) than for RF, see Figure~\ref{fig:predBR}. 



\begin{figure}[ht]
     \centering
     \label{FigRMSE}
     \begin{subfigure}[b]{0.475\textwidth}
         \centering
         \includegraphics[width=0.75\textwidth, viewport = 20 10 370 272, clip]{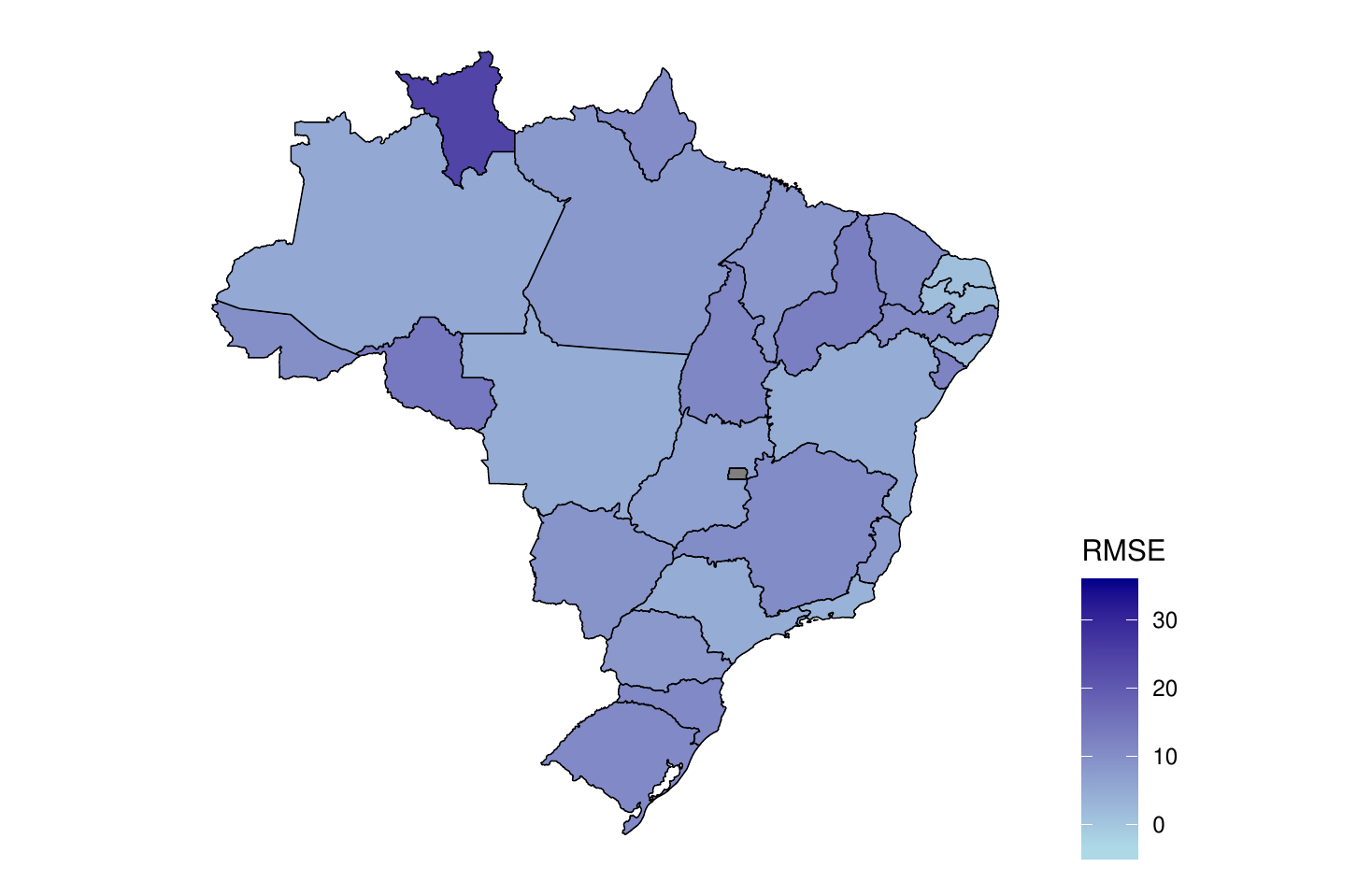}
         \caption{Random forest}
        \label{fig:RFE}
     \end{subfigure}
     \begin{subfigure}[b]{0.475\textwidth}
         \centering
         \includegraphics[width=0.75\textwidth, viewport = 20 10 370 268, clip]{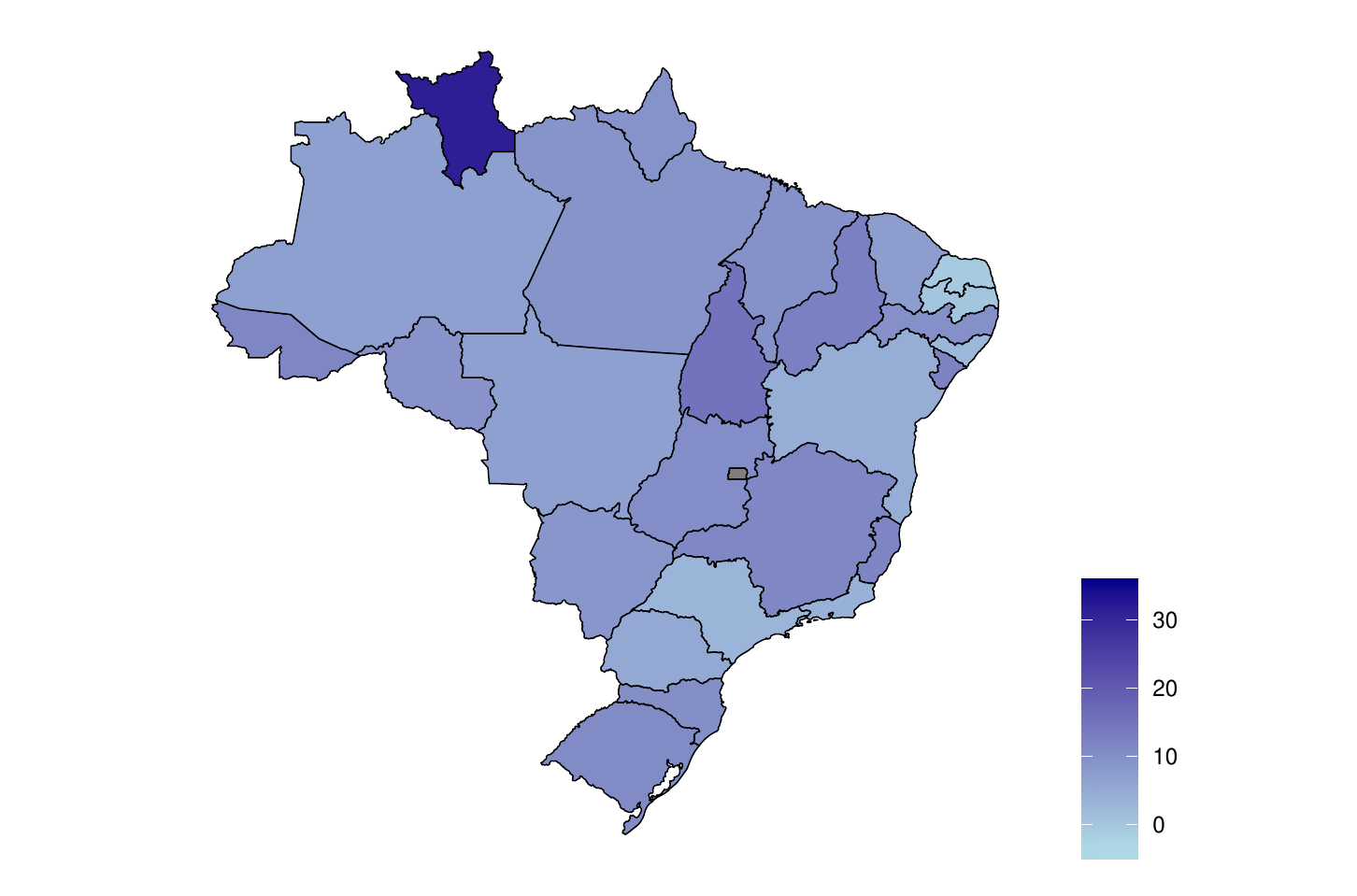}
         \caption{Boosted regression}
         \label{fig:GBME}
     \end{subfigure}
   \vskip\baselineskip
  \hfill
     \begin{subfigure}[b]{0.475\textwidth}
         \centering
         \includegraphics[width=0.75\textwidth, viewport = 20 10 370 268, clip]{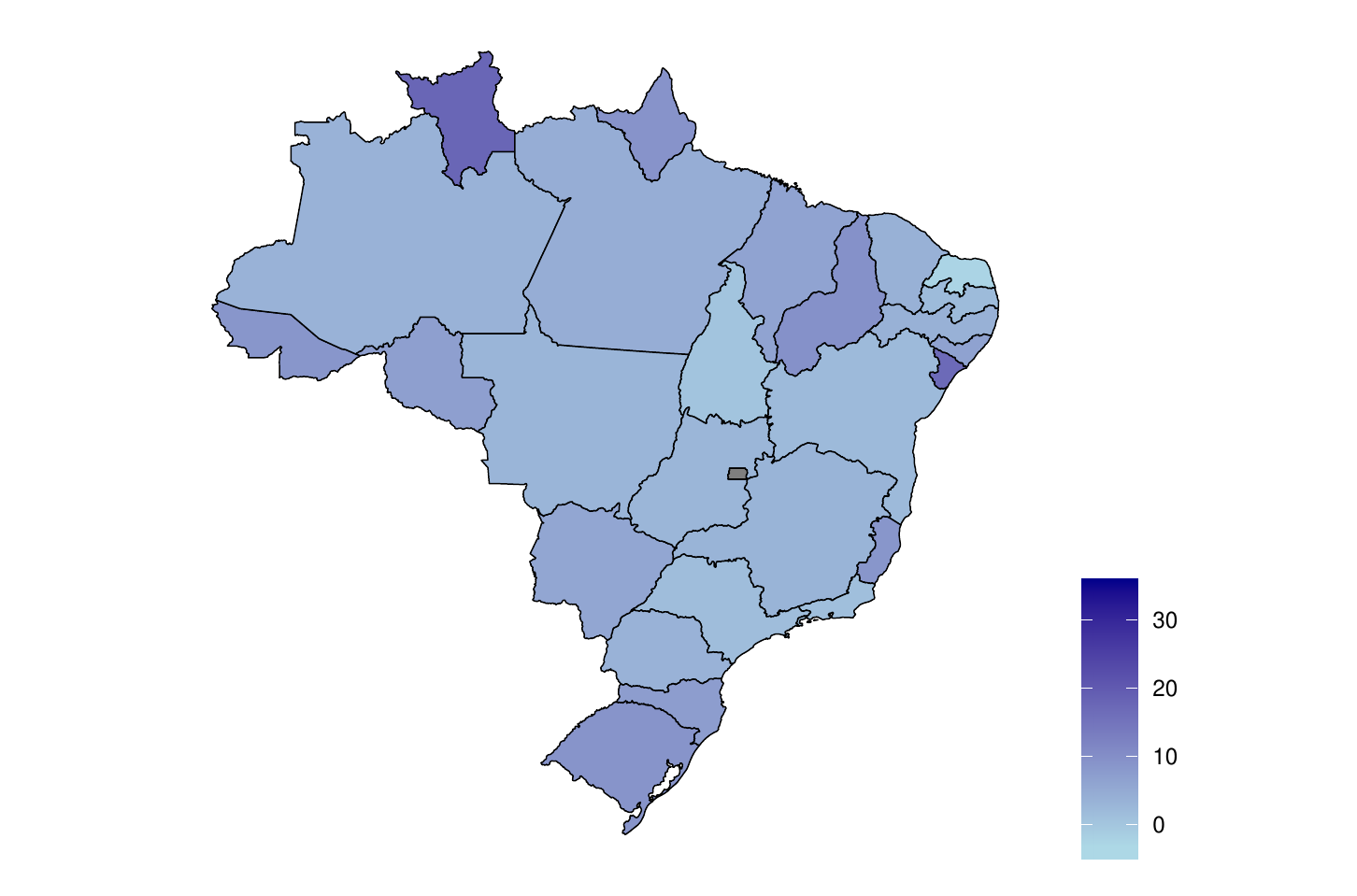}
         \caption{DFFN}
          \label{fig:DLE}
     \end{subfigure}
     \hfill
     \begin{subfigure}[b]{0.475\textwidth}
         \centering
         \includegraphics[width=0.75\textwidth, viewport = 20 10 370 268, clip]{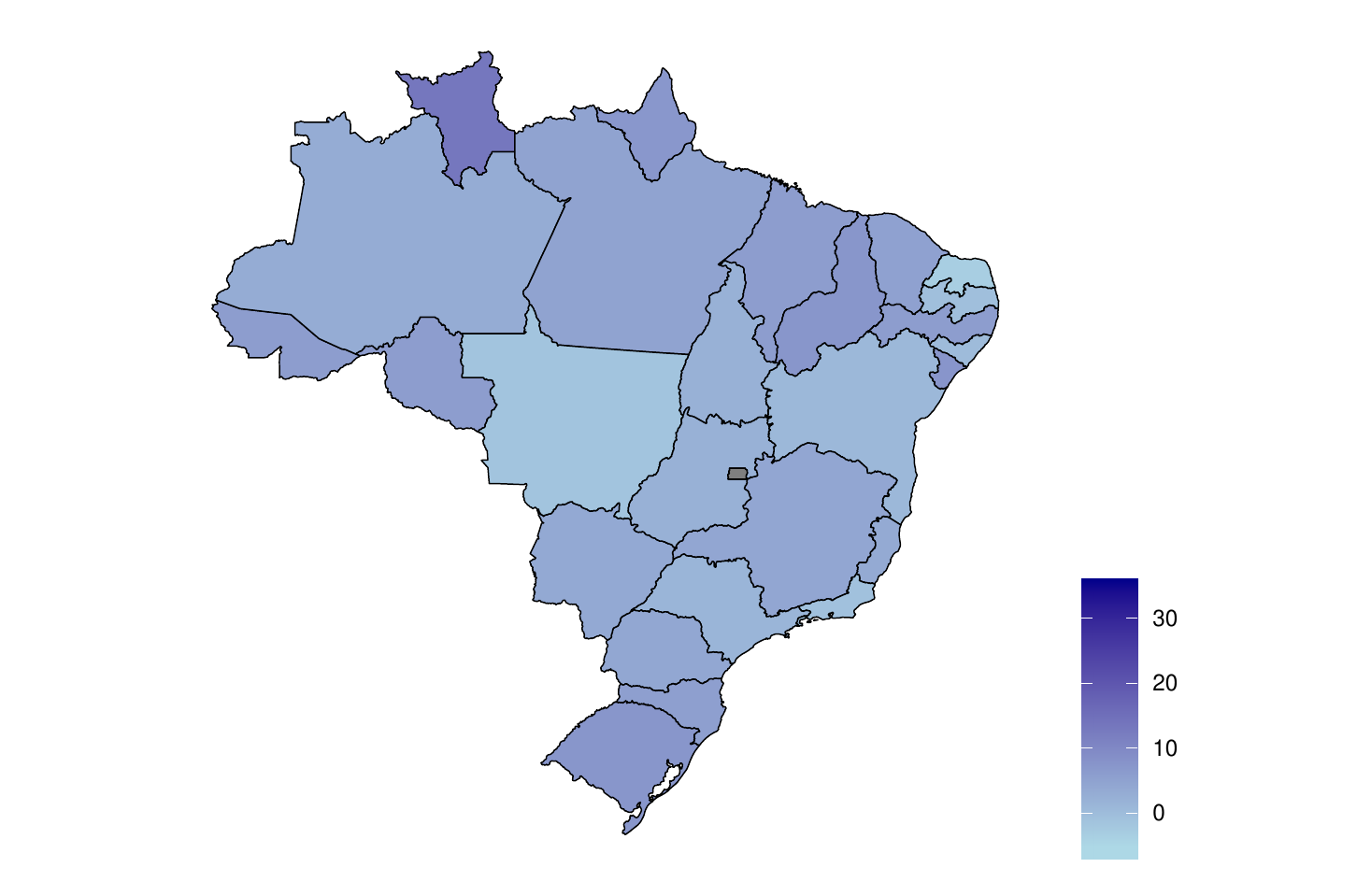}
         \caption{BMA-RMSE}
         \label{fig:BMAE}
     \end{subfigure}
        \caption{\Large Map of predictive root mean square error (RMSE) in Brazil delivered by each of the four models for Zika rate in 2018.
        }   
        \label{fig:map2}
\end{figure}

\begin{figure}[ht]
     \centering
     \begin{subfigure}[b]{0.475\textwidth}
         \centering
         \includegraphics[width=0.75\textwidth, viewport = 20 10 395 268, clip]{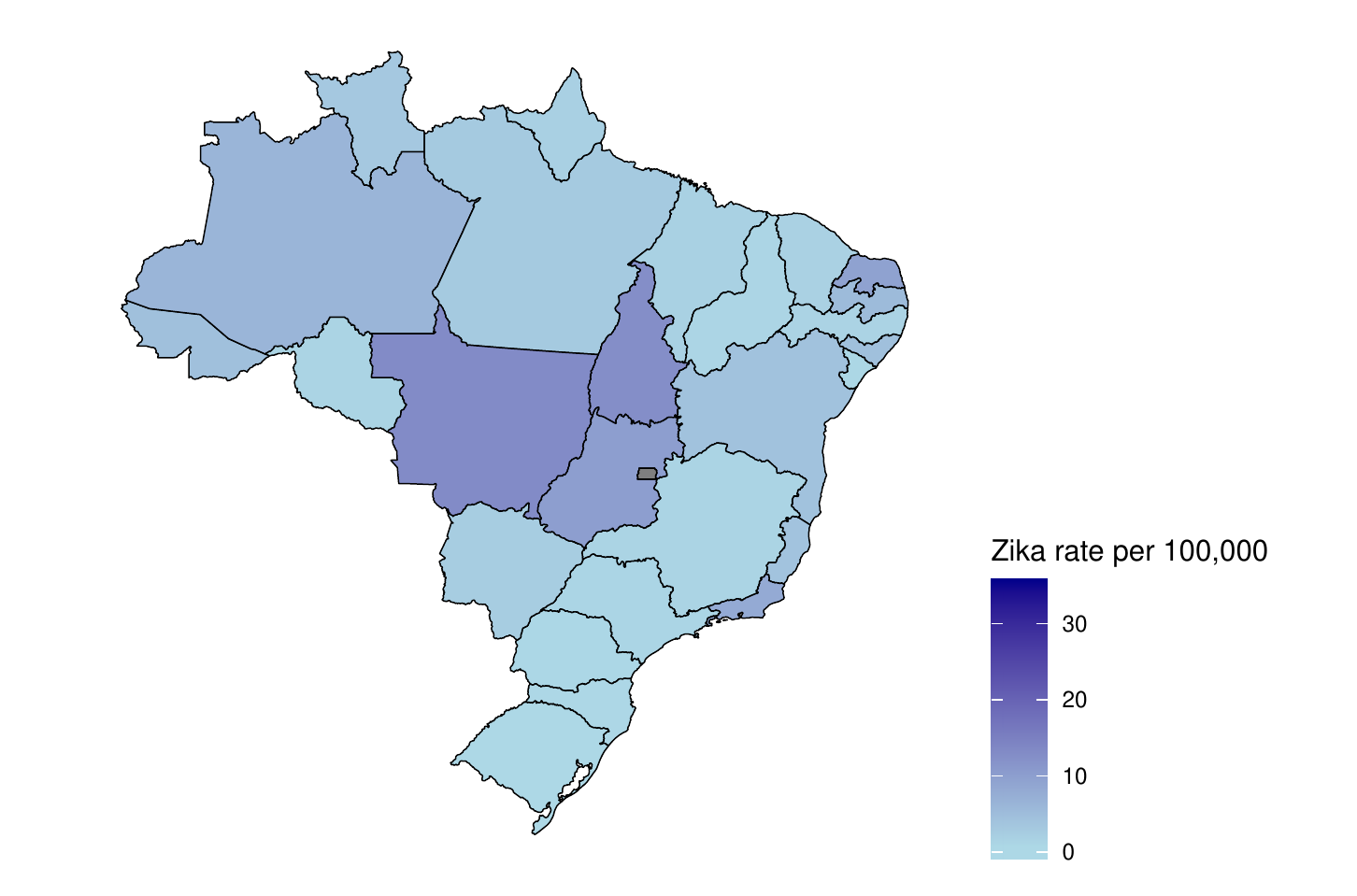}
         \caption{Observed}
        \label{fig:mapZR}
     \end{subfigure}
     \begin{subfigure}[b]{0.475\textwidth}
         \centering
         \includegraphics[width=0.7\textwidth, viewport = 20 10 375 268, clip]{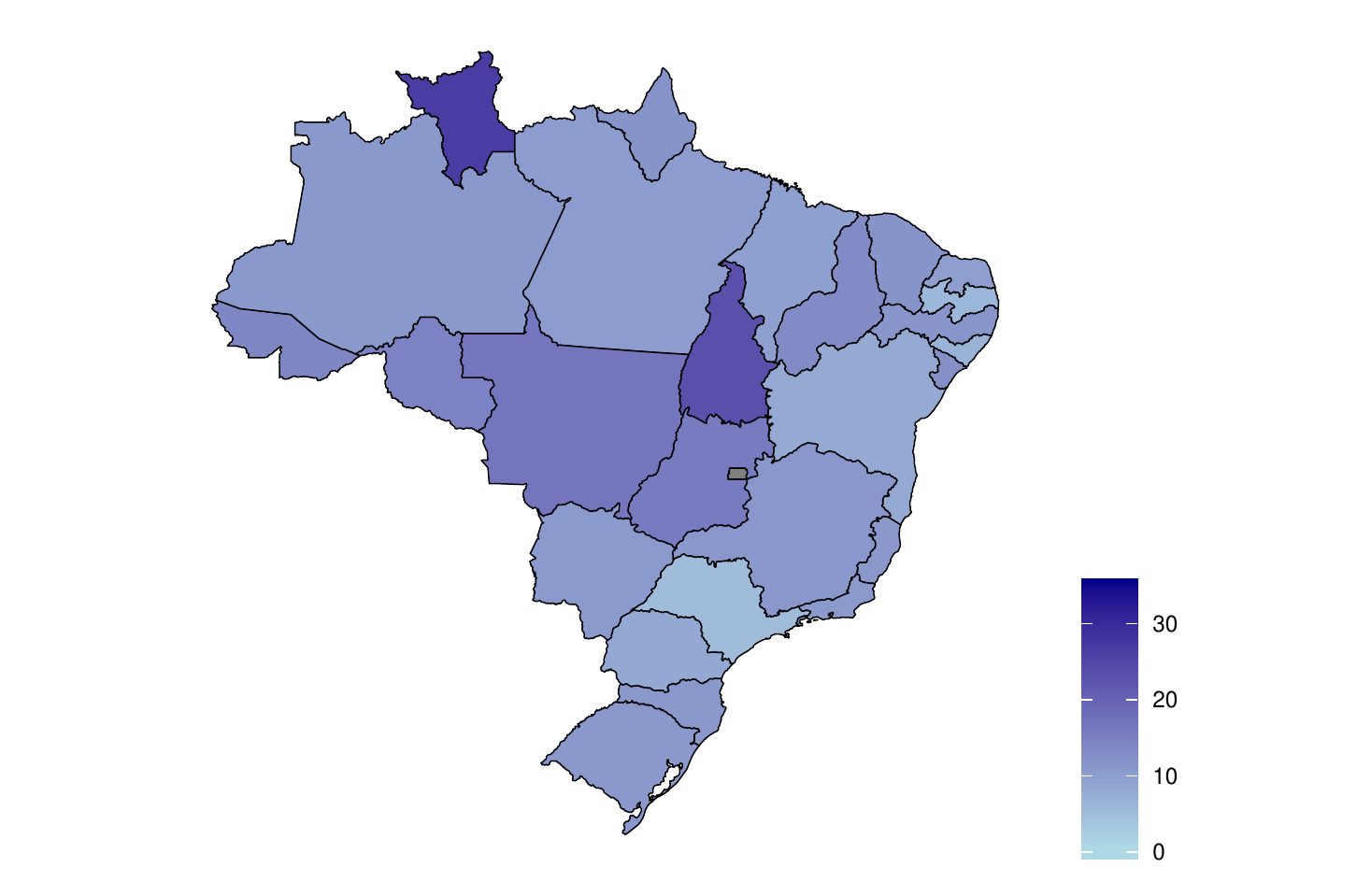}
         \caption{Random forest}
         \label{fig:mapRF}
     \end{subfigure}
   \vskip\baselineskip
  \hfill
     \begin{subfigure}[b]{0.475\textwidth}
         \centering
         \includegraphics[width=0.7\textwidth, viewport = 20 10 370 268, clip]{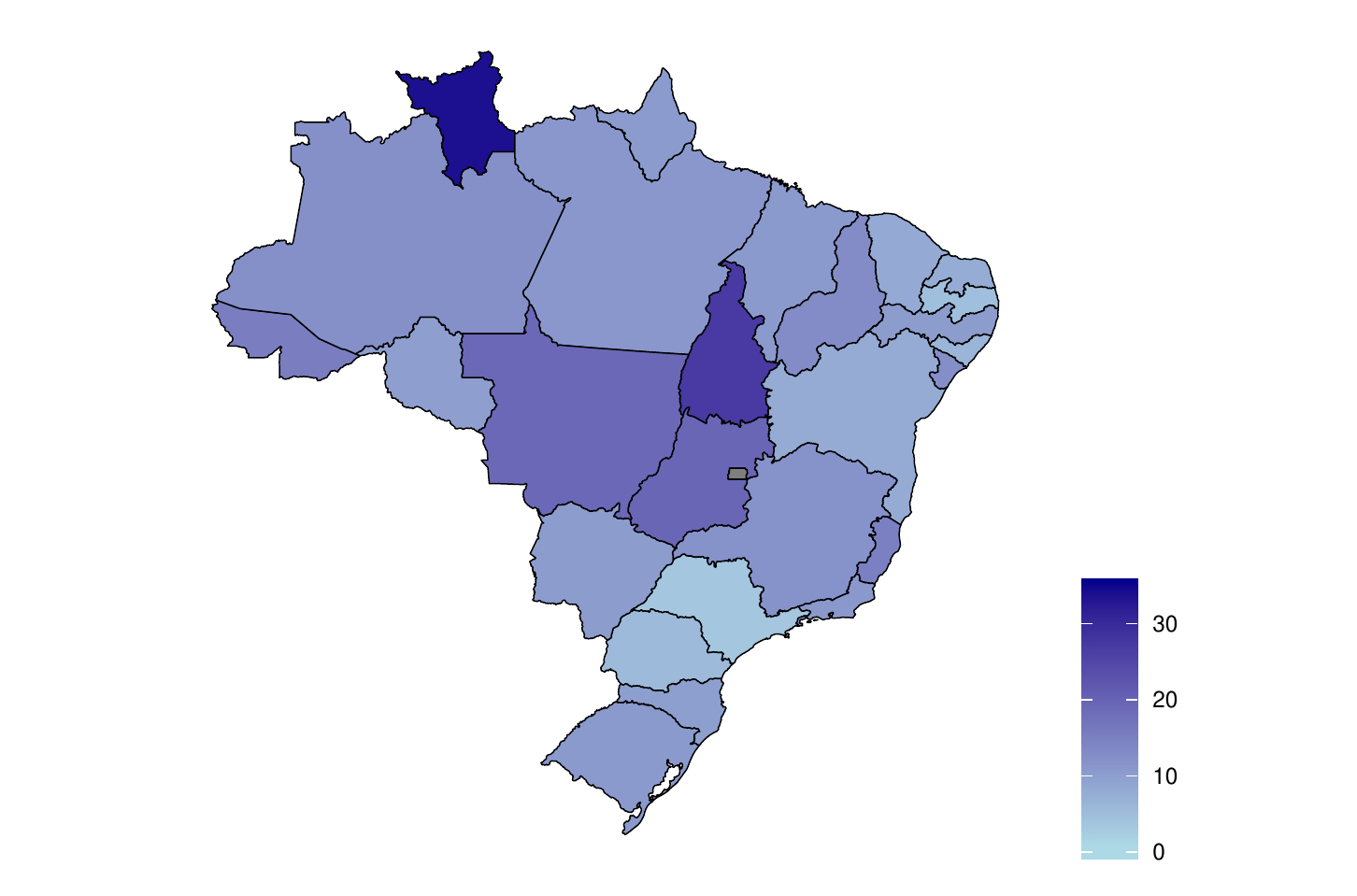}
         \caption{Boosted regression}
          \label{fig:mapGBM}
     \end{subfigure}
     \hfill
     \begin{subfigure}[b]{0.475\textwidth}
         \centering
         \includegraphics[width=0.7\textwidth, viewport = 20 10 370 268, clip]{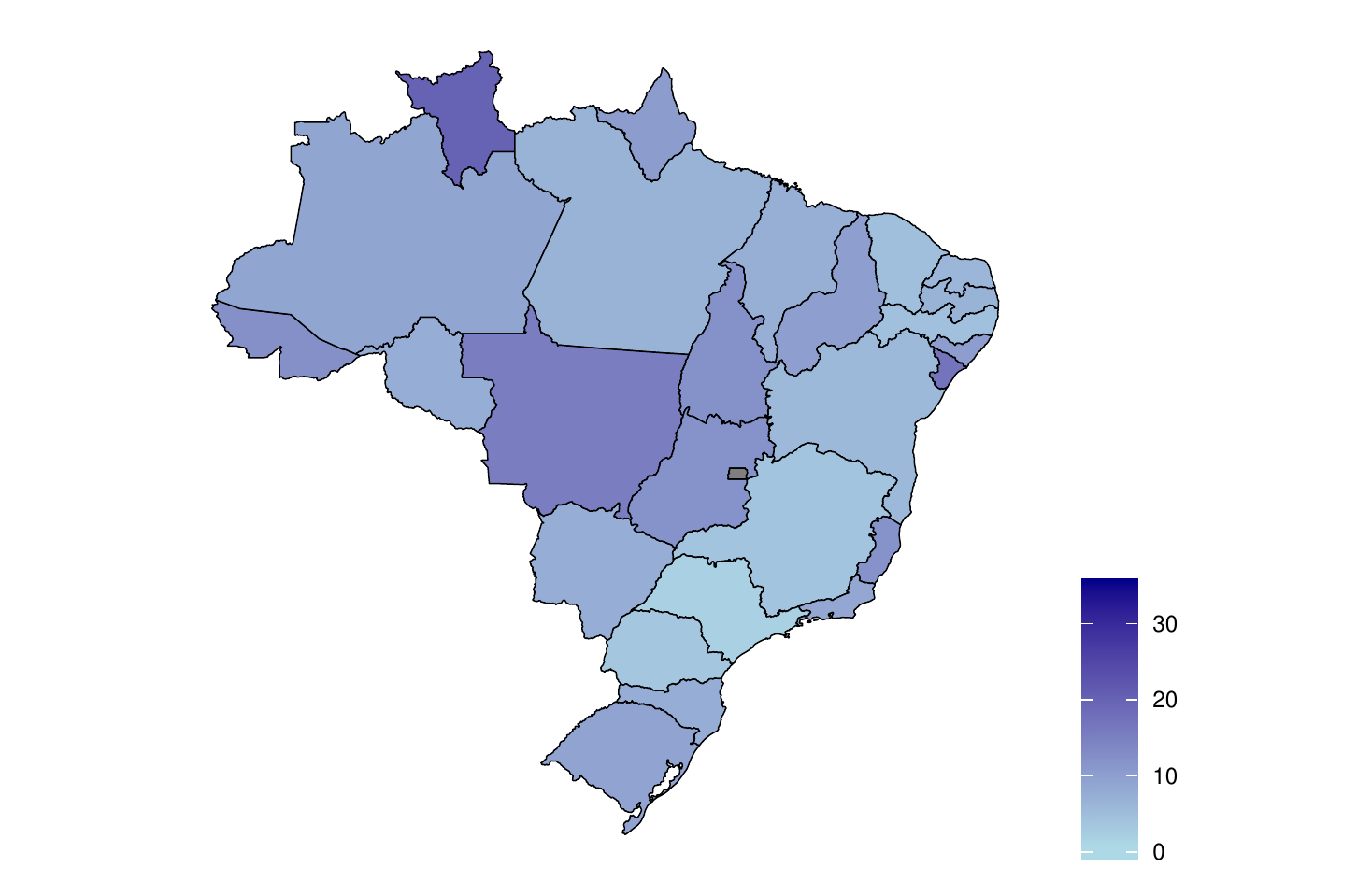}
         \caption{DFFN}
         \label{fig:mapDL}
     \end{subfigure}
     \hfill
     \begin{subfigure}[b]{0.475\textwidth}
         \centering
         \includegraphics[width=0.7\textwidth, viewport = 20 10 370 268, clip]{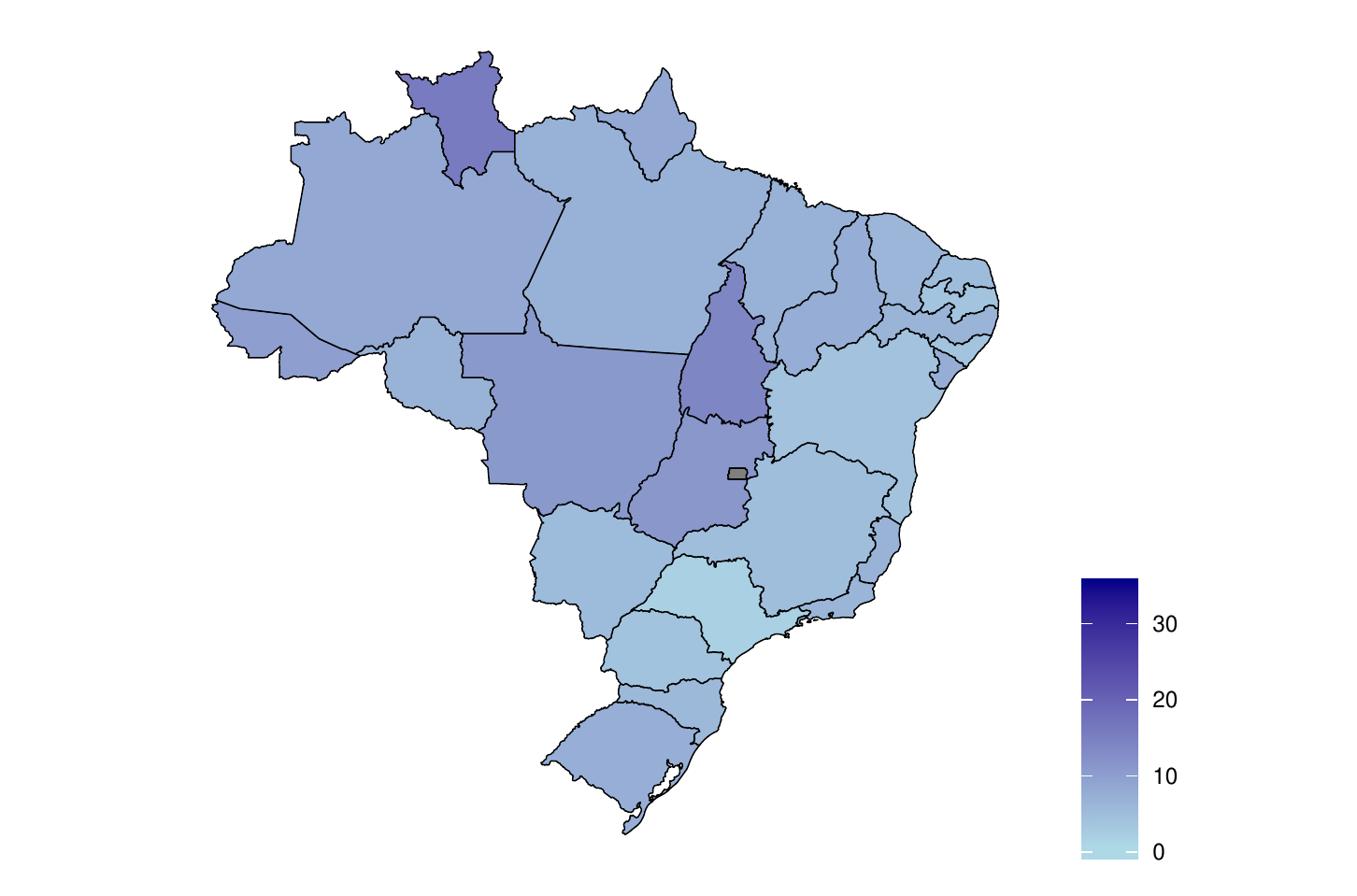}
         \caption{BMA-RMSE}
         \label{fig:mapBMA}
     \end{subfigure}
        \caption{\Large Map of the (a) observed average Zika rate in 2018 and (b--e) average predicted Zika rates in 2018 by each of the four models. 
        }   
        \label{fig:map}
\end{figure}

In addition, we examine the geographical distribution of prediction errors of each model. 
The map in Figure~\ref{fig:map2} suggests that the BMA-RMSE approach delivers the highest predictive performance for all considered states. 
Remarkably, among the best predicted states using the BMA-RMSE approach are the states with the highest infection rates, that is, Mato Grosso, Goias, and Mato Grosso do Sul~\citep[see also Figure~\ref{fig:map} showing the average Zika rates, and][]{ragas2018revisiting}.


%

{\bf Remark} Finally, while Pearson correlations reported in Table~\ref{tableRMSE_Welch} are generally low,
the resulting correlations  appear to be overall on par with the previous studies on zika in other Latin and Southern American countries~\citep{mcgough2017forecasting}.
In particular,~\cite{mcgough2017forecasting} report Pearson correlations for 3-week ahead forecast of zika in Honduras not exceeding 0.35 and as low as 0.08. Similar magnitude of Pearson correlations of about 0.30 are found for autoregressive models for 3-week ahead forecast in Colombia, while correlations in El Salvador, Martinique, and Venezuela tend to be somewhat higher. Overall, correlations reported by~\cite{mcgough2017forecasting}  tend to vary substantially among countries and among predictive models and to decrease noticeably with the increase of the forecasting horizon.

\section{Discussion}
\label{Discussion}

Zika continues to be one of the primary healthcare concerns in the Americas, Oceania, Africa and many other parts of the world -- making the virus a global challenge for healthcare professional worldwide and demanding
the world's collective preparedness for this climate-sensitive (re)emerging disease.

In this paper we have brought the concepts of topological data analysis of atmospheric variables to enhance prediction of Zika virus. In particular, we have integrated the cumulative Betti numbers as topological descriptors of 
precipitation and temperature dynamics---one of the key environmental factors in Zika spread---into three predictive machine learning models for Zika: random forest, boosted regression, and deep feed-forward neural network.
To better account for various sources of 
uncertainties and harness the power of individual predictive models, we have 
combined the resulting individual forecasts into an ensemble of the Zika spread predictions using Bayesian model averaging. Our findings based on the analysis of Zika space-time spread in Brazil have indicated that topological summaries of precipitation dynamics and air temperature contain important predictive information for the future Zika spread.

In the future we plan to advance the proposed topological approach to analysis of other related climate-sensitive diseases such as chikungunya and dengue, not only in Brazil, but also in other countries. Furthermore, epidemiological forecasting can benefit from using the tools of topological data analysis for understanding the spread of diseases through examination of the joint dynamics of topological summaries of the disease rates and associated environmental and socio-economic factors.
 
Finally, there also exist multiple directions how to better account for various types of uncertainties associated with biosurveillance of emerging climate-sensitive infectious diseases. To address the issue of limited data records and noisy epidemiological information, we plan to integrate various nontraditional data sources, such as webqueries, into surveillance and forecasting of the emerging infectious diseases. Understanding topological summaries of such nontraditional data sources and matching topological patterns of the conventional data from public health units and webqueries can shed a light on future spatial dynamics of the emerging infectious disease.  We then plan to employ the semi-parametric bootstrap to develop probabilistic forecasts associated with each model and multiple data sources,  which in return can be combined into a joint biosurveillance ensemble.

\bibliographystyle{wb_env}
\bibliography{refsZika}

\begin{thebibliography}{62}
\providecommand{\natexlab}[1]{#1}

\bibitem[{Bluman(2009)}]{bluman2009elementary}
Bluman AG, 2009. \textit{Elementary statistics: A step by step approach}.
  McGraw-Hill Higher Education New York.

\bibitem[{Breiman(1997)}]{Breiman}
Breiman L, 1997. Arcing the edge. \textit{Technical Report 486. Statistics
  Department, University of California, Berkeley} .

\bibitem[{Breiman(2001)}]{breiman2001random}
Breiman L, 2001. Random forests. \textit{Machine learning} \textbf{45}(1):
  5--32.

\bibitem[{Breiman \textit{et~al.}(2018)Breiman, Cutler, Liaw and Wiener}]{RF}
Breiman L, Cutler A, Liaw A, Wiener M, 2018. \textit{{randomForest}: {Breiman}
  and {Cutler}'s Random Forests for Classification and Regression}. {R} package
  version 4.6-14.

\bibitem[{Caminade \textit{et~al.}(2017)Caminade, Turner, Metelmann, Hesson,
  Blagrove, Solomon, Morse and Baylis}]{Caminade119}
Caminade C, Turner J, Metelmann S, Hesson JC, Blagrove MSC, Solomon T, Morse
  AP, Baylis M, 2017. Global risk model for vector-borne transmission of {Zika}
  virus reveals the role of {El Ni{\~n}o} 2015. \textit{Proceedings of the
  National Academy of Sciences} \textbf{114}(1): 119--124.

\bibitem[{Carlsson(2009)}]{carlsson2009topology}
Carlsson G, 2009. Topology and data. \textit{Bulletin of the American
  Mathematical Society} \textbf{46}(2): 255--308.

\bibitem[{Castro \textit{et~al.}(2018)Castro, Han, Carvalho, Victora and
  Fran{\c{c}}a}]{castro2018implications}
Castro MC, Han QC, Carvalho LR, Victora CG, Fran{\c{c}}a GV, 2018. Implications
  of {Zika} virus and congenital {Zika} syndrome for the number of live births
  in {Brazil}. \textit{Proceedings of the National Academy of Sciences}
  \textbf{115}(24): 6177--6182.

\bibitem[{{CDC}(2018)}]{WinNT3}
{CDC}, 2018. Centers for disease control and prevention ({About Zika}).

\bibitem[{Chazal and Michel(2017)}]{chazal2017introduction}
Chazal F, Michel B, 2017. An introduction to topological data analysis:
  Fundamental and practical aspects for data scientists. \textit{arXiv Preprint
  arXiv:1710.04019} .

\bibitem[{Costa and {\v{S}}kraba(2014)}]{costa2014topological}
Costa JP, {\v{S}}kraba P, 2014. A topological data analysis approach to
  epidemiology. In \textit{European Conference of Complexity Science}.

\bibitem[{Dick(1952)}]{dick1952zikaII}
Dick G, 1952. Zika virus ({II}). {Pathogenicity} and physical properties.
  \textit{Transactions of the Royal Society of Tropical Medicine and Hygiene}
  \textbf{46}(5): 521--534.

\bibitem[{Dick \textit{et~al.}(1952)Dick, Kitchen and Haddow}]{dick1952zika}
Dick G, Kitchen S, Haddow A, 1952. Zika virus ({I}). {Isolations} and
  serological specificity. \textit{Transactions of the Royal Society of
  Tropical Medicine and Hygiene} \textbf{46}(5): 509--520.

\bibitem[{Diggle \textit{et~al.}(2005)Diggle, Rowlingson and
  Su}]{diggle2005point}
Diggle P, Rowlingson B, Su Tl, 2005. Point process methodology for on-line
  spatio-temporal disease surveillance. \textit{Environmetrics: The official
  journal of the International Environmetrics Society} \textbf{16}(5):
  423--434.

\bibitem[{Fasy \textit{et~al.}(2018)Fasy, Kim, Lecci, Maria, Millman, Rouvreau,
  Morozov, Bauer, Kerber and Reininghaus}]{TDA}
Fasy BT, Kim J, Lecci F, Maria C, Millman DL, Rouvreau V, Morozov D, Bauer U,
  Kerber M, Reininghaus J, 2018. \textit{TDA: Statistical Tools for Topological
  Data Analysis}. {R package version 1.6.4}.

\bibitem[{Ferraris \textit{et~al.}(2019)Ferraris, Yssel and
  Miss{\'e}}]{ferraris2019zika}
Ferraris P, Yssel H, Miss{\'e} D, 2019. Zika virus infection: An update.
  \textit{Microbes and infection} .

\bibitem[{Fragoso \textit{et~al.}(2018)Fragoso, Bertoli and
  Louzada}]{fragoso2018bayesian}
Fragoso TM, Bertoli W, Louzada F, 2018. Bayesian model averaging: A systematic
  review and conceptual classification. \textit{International Statistical
  Review} \textbf{86}(1): 1--28.

\bibitem[{Friedman(2001)}]{friedman2001greedy}
Friedman JH, 2001. Greedy function approximation: a gradient boosting machine.
  \textit{Annals of Statistics} \textbf{29}(5): 1189--1232.

\bibitem[{Friedman(2002)}]{friedman2002stochastic}
Friedman JH, 2002. Stochastic gradient boosting. \textit{Computational
  Statistics \& Data Analysis} \textbf{38}(4): 367--378.

\bibitem[{Gidea and Katz(2018)}]{gidea2018topological}
Gidea M, Katz Y, 2018. Topological data analysis of financial time series:
  Landscapes of crashes. \textit{Physica A: Statistical Mechanics and Its
  Applications} \textbf{491}: 820--834.

\bibitem[{Goeijenbier \textit{et~al.}(2016)Goeijenbier, Slobbe, Van~der Eijk,
  de~Mendon{\c{c}}a~Melo, Koopmans and Reusken}]{goeijenbier2016zika}
Goeijenbier M, Slobbe L, Van~der Eijk A, de~Mendon{\c{c}}a~Melo M, Koopmans M,
  Reusken C, 2016. Zika virus and the current outbreak: an overview.
  \textit{Neth J Med} \textbf{74}(3): 104--9.

\bibitem[{Greenwell \textit{et~al.}(2019)Greenwell, Boehmke, Cunningham and
  Developers}]{gbm}
Greenwell B, Boehmke B, Cunningham J, Developers G, 2019. \textit{gbm:
  Generalized Boosted Regression Models}. R package version 2.1.5.

\bibitem[{Hagan and Menhaj(1994)}]{hagan1994training}
Hagan MT, Menhaj MB, 1994. Training feedforward networks with the {Marquardt}
  algorithm. \textit{IEEE Transactions on Neural Networks} \textbf{5}(6):
  989--993.

\bibitem[{Hastie \textit{et~al.}(2009)Hastie, Tibshirani and
  Friedman}]{Hastie:etal:2009}
Hastie TJ, Tibshirani RJ, Friedman JH, 2009. \textit{The Elements of
  Statistical Learning: Data Mining, Inference, and Prediction}. Springer, New
  York, 2 edition.

\bibitem[{Heukelbach \textit{et~al.}(2016)Heukelbach, Alencar, Kelvin,
  de~Oliveira and de~G{\'o}es~Cavalcanti}]{heukelbach2016zika}
Heukelbach J, Alencar CH, Kelvin AA, de~Oliveira WK, de~G{\'o}es~Cavalcanti LP,
  2016. Zika virus outbreak in {Brazil}. \textit{The Journal of Infection in
  Developing Countries} \textbf{10}(02): 116--120.

\bibitem[{Hoeting \textit{et~al.}(1999)Hoeting, Madigan, Raftery and
  Volinsky}]{hoeting1999bayesian}
Hoeting JA, Madigan D, Raftery AE, Volinsky CT, 1999. Bayesian model averaging:
  a tutorial. \textit{Statistical Science} \textbf{14}(4): 382--401.

\bibitem[{INMET(2019)}]{INMET}
INMET, 2019. Instituto nacional de meteorologia.
  \url{http://www.inmet.gov.br/portal/index.php?r=home2/index}.

\bibitem[{Islambekov and Gel(2019)}]{islambekov2019unsupervised}
Islambekov U, Gel YR, 2019. Unsupervised space--time clustering using
  persistent homology. \textit{Environmetrics} \textbf{30}(4): e2539.

\bibitem[{Islambekov \textit{et~al.}(2019)Islambekov, Yuvaraj and
  Gel}]{islambekov2019harnessing}
Islambekov U, Yuvaraj M, Gel YR, 2019. Harnessing the power of topological data
  analysis to detect change points. \textit{Environmetrics} : e2612.

\bibitem[{Jang \textit{et~al.}(2007)Jang, Lee, Lawson and
  Browne}]{jang2007comparison}
Jang MJ, Lee Y, Lawson AB, Browne WJ, 2007. A comparison of the hierarchical
  likelihood and bayesian approaches to spatial epidemiological modelling.
  \textit{Environmetrics: The official journal of the International
  Environmetrics Society} \textbf{18}(7): 809--821.

\bibitem[{Kim and Ahn(2019)}]{kim2019weekly}
Kim J, Ahn I, 2019. Weekly {ILI} patient ratio change prediction using news
  articles with support vector machine. \textit{BMC Bioinformatics}
  \textbf{20}(1): 259.

\bibitem[{LeDell \textit{et~al.}(2019)LeDell, Gill, Aiello, Fu, Candel, Click,
  Kraljevic, Nykodym, Aboyoun, Kurka and Malohlava}]{h2o}
LeDell E, Gill N, Aiello S, Fu A, Candel A, Click C, Kraljevic T, Nykodym T,
  Aboyoun P, Kurka M, Malohlava M, 2019. \textit{h2o: R Interface for `H2O'}. R
  package version 3.26.0.2.

\bibitem[{Li \textit{et~al.}(2020)Li, Ofori-Boateng, Gel and
  Zhang}]{li2020hybrid}
Li B, Ofori-Boateng D, Gel YR, Zhang J, 2020. A hybrid approach for
  transmission grid resilience assessment using reliability metrics and power
  system local network topology. \textit{Sustainable and Resilient
  Infrastructure} : 1--16.

\bibitem[{Lo and Park(2018)}]{lo2018modeling}
Lo D, Park B, 2018. Modeling the spread of the {Zika} virus using topological
  data analysis. \textit{PloS One} \textbf{13}(2): e0192120.

\bibitem[{Loh(2011)}]{loh2011k}
Loh JM, 2011. K-scan for anomaly detection in disease surveillance.
  \textit{Environmetrics} \textbf{22}(2): 179--191.

\bibitem[{Malone \textit{et~al.}(2016)Malone, Homan, Callahan,
  Glasspool-Malone, Damodaran, Schneider, Zimler, Talton, Cobb, Ruzic
  \textit{et~al.}}]{malone2016zika}
Malone RW, Homan J, Callahan MV, Glasspool-Malone J, Damodaran L, Schneider
  ADB, Zimler R, Talton J, Cobb RR, Ruzic I, \textit{et~al.}, 2016. Zika virus:
  medical countermeasure development challenges. \textit{PLoS Neglected
  Tropical Diseases} \textbf{10}(3): e0004530.

\bibitem[{Manore and Hyman(2016)}]{manore2016mathematical}
Manore C, Hyman M, 2016. Mathematical models for fighting {Zika} virus.
  \textit{Siam News} .

\bibitem[{McDermott and Wikle(2019)}]{mcdermott2019deep}
McDermott PL, Wikle CK, 2019. Deep echo state networks with uncertainty
  quantification for spatio-temporal forecasting. \textit{Environmetrics}
  \textbf{30}(3): e2553.

\bibitem[{McGough \textit{et~al.}(2017)McGough, Brownstein, Hawkins and
  Santillana}]{mcgough2017forecasting}
McGough SF, Brownstein JS, Hawkins JB, Santillana M, 2017. Forecasting {Zika}
  incidence in the 2016 {Latin America} outbreak combining traditional disease
  surveillance with search, social media, and news report data. \textit{PLoS
  Neglected Tropical Diseases} \textbf{11}(1): e0005295.

\bibitem[{MHB(2018)}]{WinNT5}
MHB, 2018. {Ministry of Health of Brazil}.
  \url{http://portalms.saude.gov.br/boletins-epidemiologicos}.

\bibitem[{Mu{\~n}oz \textit{et~al.}(2017)Mu{\~n}oz, Thomson, Stewart-Ibarra,
  Vecchi, Chourio, N{\'a}jera, Moran and Yang}]{munoz2017could}
Mu{\~n}oz {\'A}G, Thomson MC, Stewart-Ibarra AM, Vecchi GA, Chourio X,
  N{\'a}jera P, Moran Z, Yang X, 2017. Could the recent {Zika} epidemic have
  been predicted? \textit{Frontiers in Microbiology} \textbf{8}: 1291.

\bibitem[{Nobre \textit{et~al.}(2005)Nobre, Schmidt and
  Lopes}]{nobre2005spatio}
Nobre AA, Schmidt AM, Lopes HF, 2005. Spatio-temporal models for mapping the
  incidence of malaria in par{\'a}. \textit{Environmetrics} \textbf{16}(3):
  291--304.

\bibitem[{Peck \textit{et~al.}(2015)Peck, Olsen and
  Devore}]{peck2015introduction}
Peck R, Olsen C, Devore JL, 2015. \textit{Introduction to statistics and data
  analysis}. Cengage Learning.

\bibitem[{Raftery \textit{et~al.}(2005)Raftery, Gneiting, Balabdaoui and
  Polakowski}]{raftery2005using}
Raftery AE, Gneiting T, Balabdaoui F, Polakowski M, 2005. Using {Bayesian}
  model averaging to calibrate forecast ensembles. \textit{Monthly Weather
  Review} \textbf{133}(5): 1155--1174.

\bibitem[{Ragas(2018)}]{ragas2018revisiting}
Ragas J, 2018. Revisiting global health from the periphery: the {Zika} virus.
  \textit{Hist{\'o}ria, Ci{\^e}ncias, Sa{\'u}de-Manguinhos} \textbf{25}(4):
  1185--1187.

\bibitem[{Rees \textit{et~al.}(2018)Rees, Petukhova, Mascarenhas, Pelcat and
  Ogden}]{rees2018environmental}
Rees EE, Petukhova T, Mascarenhas M, Pelcat Y, Ogden NH, 2018. Environmental
  and social determinants of population vulnerability to {Zika} virus emergence
  at the local scale. \textit{Parasites \& Vectors} \textbf{11}(1): 290.

\bibitem[{Ridgeway(2019)}]{ridgeway2007generalized}
Ridgeway G, 2019. Generalized boosted models: A guide to the gbm package.
  \textit{Update} \textbf{1}(1): 2019.

\bibitem[{Riedmiller and Braun(1993)}]{riedmiller1993direct}
Riedmiller M, Braun H, 1993. A direct adaptive method for faster
  backpropagation learning: The {RPROP} algorithm. In \textit{Proceedings of
  the IEEE International Conference on Neural Networks}, volume 1993, San
  Francisco, CA, 586--591.

\bibitem[{Saggar \textit{et~al.}(2018)Saggar, Sporns, Gonzalez-Castillo,
  Bandettini, Carlsson, Glover and Reiss}]{saggar2018towards}
Saggar M, Sporns O, Gonzalez-Castillo J, Bandettini PA, Carlsson G, Glover G,
  Reiss AL, 2018. Towards a new approach to reveal dynamical organization of
  the brain using topological data analysis. \textit{Nature Communications}
  \textbf{9}(1): 1399.

\bibitem[{Self \textit{et~al.}(2018)Self, McMahan, Brown, Lund, Gettings and
  Yabsley}]{self2018large}
Self SCW, McMahan CS, Brown DA, Lund RB, Gettings JR, Yabsley MJ, 2018. A
  large-scale spatio-temporal binomial regression model for estimating
  seroprevalence trends. \textit{Environmetrics} \textbf{29}(8): e2538.

\bibitem[{Seo \textit{et~al.}(2018)Seo, Lee and Kim}]{seo2018decoding}
Seo J, Lee J, Kim K, 2018. Decoding of polar code by using deep feed-forward
  neural networks. In \textit{2018 International Conference on Computing,
  Networking and Communications (ICNC)}, IEEE, 238--242.

\bibitem[{Soliman \textit{et~al.}(2019)Soliman, Lyubchich and
  Gel}]{soliman2019complementing}
Soliman M, Lyubchich V, Gel YR, 2019. Complementing the power of deep learning
  with statistical model fusion: Probabilistic forecasting of influenza in
  {Dallas County, Texas, USA}. \textit{Epidemics} \textbf{28}: 100345.

\bibitem[{Starnes \textit{et~al.}(2010)Starnes, Yates and
  Moore}]{starnes2010practice}
Starnes DS, Yates D, Moore DS, 2010. \textit{The practice of statistics}.
  Macmillan.

\bibitem[{Suparit \textit{et~al.}(2018)Suparit, Wiratsudakul and
  Modchang}]{suparit2018mathematical}
Suparit P, Wiratsudakul A, Modchang C, 2018. A mathematical model for {Zika}
  virus transmission dynamics with a time-dependent mosquito biting rate.
  \textit{Theoretical Biology and Medical Modelling} \textbf{15}(1): 11.

\bibitem[{Teng \textit{et~al.}(2017)Teng, Bi, Xie, Jin, Huang, Lin, An, Feng
  and Tong}]{teng2017dynamic}
Teng Y, Bi D, Xie G, Jin Y, Huang Y, Lin B, An X, Feng D, Tong Y, 2017. Dynamic
  forecasting of {Zika} epidemics using {Google Trends}. \textit{PloS One}
  \textbf{12}(1): e0165085.

\bibitem[{Tjaden \textit{et~al.}(2013)Tjaden, Thomas, Fischer and
  Beierkuhnlein}]{tjaden2013extrinsic}
Tjaden NB, Thomas SM, Fischer D, Beierkuhnlein C, 2013. Extrinsic incubation
  period of dengue: Knowledge, backlog, and applications of temperature
  dependence. \textit{PLoS Neglected Tropical Diseases} \textbf{7}(6): e2207.

\bibitem[{Torabi(2013)}]{torabi2013spatio}
Torabi M, 2013. Spatio--temporal modeling for disease mapping using car and
  b-spline smoothing. \textit{Environmetrics} \textbf{24}(3): 180--188.

\bibitem[{Ugarte \textit{et~al.}(2009)Ugarte, Goicoa, Ibanez and
  Militino}]{ugarte2009evaluating}
Ugarte M, Goicoa T, Ibanez B, Militino A, 2009. Evaluating the performance of
  spatio-temporal bayesian models in disease mapping. \textit{Environmetrics:
  The official journal of the International Environmetrics Society}
  \textbf{20}(6): 647--665.

\bibitem[{Welch(1947)}]{welch1947generalization}
Welch BL, 1947. The generalization of student's' problem when several different
  population variances are involved. \textit{Biometrika} \textbf{34}(1/2):
  28--35.

\bibitem[{{WHO}(2019)}]{WinNT4}
{WHO}, 2019. {World Health Organization}.
  \url{http://www.who.int/emergencies/diseases/zika/en/}.

\bibitem[{World-Weather-Online(2018)}]{WinNT6}
World-Weather-Online, 2018. \url{www.worldweatheronline.com/}.

\bibitem[{Zhang \textit{et~al.}(2007)Zhang, Zhang, Lok and
  Lyu}]{zhang2007hybrid}
Zhang JR, Zhang J, Lok TM, Lyu MR, 2007. A hybrid particle swarm
  optimization--back-propagation algorithm for feedforward neural network
  training. \textit{Applied Mathematics and Computation} \textbf{185}(2):
  1026--1037.

\bibitem[{Zomorodian and Carlsson(2005)}]{zomorodian2005computing}
Zomorodian A, Carlsson G, 2005. Computing persistent homology. \textit{Discrete
  \& Computational Geometry} \textbf{33}(2): 249--274.

\end{thebibliography}

\end{document}